\documentclass[manuscript]{aastex}

\newcommand{\kms}{\mbox{km\,s$^{-1}$}}

\slugcomment{v.5, To appear in PASA}

\shorttitle{The Southern 2MASS AGN Survey}
\shortauthors{Masci et al.}

\begin{document}


\title{The Southern 2MASS AGN Survey: spectroscopic\\ follow-up with 6dF}


\author{Frank J. Masci and Roc M. Cutri}
\affil{Infrared Processing and Analysis Center, Caltech 100-22,
     Pasadena, CA 91125, USA}
\email{fmasci@caltech.edu}

\author{Paul J. Francis}
\affil{Australian National University, ACT 0200, Australia}

\author{Brant O. Nelson\altaffilmark{1}}
\affil{Infrared Processing and Analysis Center, Caltech 100-22,
     Pasadena, CA 91125, USA}
     
\author{John P. Huchra}
\affil{Harvard-Smithsonian Center for Astrophysics,
     Cambridge, MA 02138, USA}

\and
     
\author{D. Heath Jones, Matthew Colless, and Will Saunders}
\affil{Anglo-Australian Observatory, P.O. Box 296,
     Epping, NSW 1710, Australia}


\altaffiltext{1}{present address: Vermont Academy, Saxtons River,
                 VT 05154, USA}


\begin{abstract}
The Two Micron All-Sky Survey (2MASS) has provided a uniform photometric
catalog to search for previously unknown red AGN and QSOs.
We have extended the search to the southern equatorial sky 
by obtaining spectra for 1182 AGN candidates using the Six Degree Field (6dF)
multifibre spectrograph on the UK Schmidt Telescope. These were scheduled as
auxiliary targets for the 6dF Galaxy Redshift Survey.
The candidates were selected using a single color
cut of $J - K_s > 2$ to $K_s\,\lesssim\,15.5$ and a galactic latitude
of $|b|>30^\circ$. 432 spectra were of sufficient quality to enable a reliable 
classification. 116 sources (or $\simeq$ 27\%) were securely classified as type
1 AGN, 20 as probable type 1s, and 57 as probable type 2 AGN.  
Most of them span the redshift range $0.05<z<0.5$
and only 8 (or $\sim6\%$) were previously identified as AGN or QSOs.
Our selection leads to a significantly higher AGN 
identification rate amongst local galaxies ($>20\%$) than in  
any previous (mostly blue-selected) galaxy survey.
A small fraction of the type 1 AGN could have their optical colors
reddened by optically thin dust with $A_V<2$ mag relative to optically
selected QSOs. A handful show evidence for excess far-IR emission.
The equivalent width (EW) and color distributions
of the type 1 and 2 AGN are consistent with AGN unified models.
In particular, the EW of the [O {\scriptsize III}] emission line
weakly correlates with optical--near-IR color in each class of AGN,
suggesting anisotropic obscuration of the AGN continuum.
Overall, the optical properties of the 2MASS red AGN are not
dramatically different from those of optically-selected QSOs.
Our near-IR selection appears to detect the most near-IR
luminous QSOs in the local universe to $z\simeq0.6$
and provides incentive to extend the search to deeper
near-IR surveys. 
\end{abstract}





\section{Introduction}

Much of our knowledge about the distribution and properties
of Active Galactic Nuclei (AGN) has come from 
samples that are flux-limited at blue optical wavelengths. This is because
their spectral energy distributions (SEDs) generally exhibit
a UV flux excess. Such surveys
can be very efficient and complete down to their blue 
flux limit, for example, the
Large Bright Quasar Survey \citep[LBQS;][]{22}.
However, any survey with a blue-flux limit will be relatively biased 
against objects whose intrinsic emission peaks at other wavelengths, 
e.g., near-IR emission from the host galaxy 
\citep[][]{40,2}.
Alternatively, the optical/UV can be partially or heavily absorbed by dust,
as revealed by radio, near-to-mid infrared, and X-ray surveys
\citep[][]{60,18,42,9}.
The Sloan Digital Sky Survey \citep[SDSS;][]{43} has employed
a range of multicolor optical selection techniques with relaxed constraints on
morphology to search for QSOs to redshifts of $\simeq6$. This has reduced the 
bias relative to the simple UV-excess
criteria used in early surveys, but has not eliminated it. 

A complete census of AGN over cosmic time is essential for understanding 
galaxy formation and evolution since the properties of the central
black hole and host galaxy are intimately linked \citep[see reviews in][]{24}.
The fraction of AGN missing from optically selected samples,
both as a function of redshift and luminosity
is an important parameter. This is expected to be related to
the duty cycle of black-hole fueling and timescales for 
regulating star formation. Values are currently very uncertain and
range from 15 to over 50\% \citep[e.g.,][]{44,4,16,17}.
This uncertainty is due to selection of the appropriate comparison sample 
of optically selected AGN, properties of the ``unbiased'' sample
and how representative it is, and
difficulties in quantifying the amount
of bias (e.g., dust extinction).

\subsection{Near-IR Selection}

\citet{58} showed that it is possible to construct
a complete $K$-band--limited QSO sample by combining optical and near-IR
photometry. This exploits the characteristic K-band excess seen in 
QSOs compared to stars and has been termed the KX-selection method. 
A number of small pilot surveys have employed KX-selection 
and variants thereof \citep[e.g.,][]{6,50,31,52}. The largest
is being compiled from the UKIDSS Large Area Survey covering 
~12.8 deg$^2$ \citep[][]{37}. This survey currently reports a surface 
density of $\simeq15$ deg$^{-2}$ for broad-lined AGN
with $K\leq17$ and $z<3$. They estimated that about 50\% 
were missed by SDSS, therefore revealing a large population of red QSOs.

Although the KX method is turning
up many luminous red QSOs to high redshift, the situation is 
different for less luminous local AGN (e.g., Seyfert nuclei). These  
are usually not found by color selection since host galaxy light can
dominate their broadband colors. Historically, they are found by taking 
optical spectra of the nuclear regions of large samples of galaxies
\citep[e.g.,][]{23}. Many of these samples are
flux-limited at blue optical wavelengths and hence subject to bias.
The most significant bias is that the blue light
is dominated by recent star formation
which overwhelms any emission from a central AGN.
A near-IR selected sample of local galaxies using color criteria 
analogous to that used in KX methods will reduce this bias (see below).

The largest near-IR survey to date is the Two Micron All Sky Survey
\citep[2MASS;][]{51}. There have been several studies that searched 
for AGN in 2MASS. The first was that from \citet{7} in
the northern equatorial sky. They selected
sources with $J - K_s > 2$ and $K_s\leq15.5$\footnote{$K_s$ is similar to the
$K$ filter but cutting off at a shorter red wavelength to minimize thermal
emission.} at galactic latitude $|b|>30\deg$.
Spectroscopic follow-up revealed that $\sim75\%$ were previously
unidentified AGN, with $\sim80\%$ of these associated with broad 
emission-line (type 1) AGN, i.e., Seyfert 1s and QSOs, and the remainder were 
narrow-line (type 2) AGN, typically Seyfert 2s, type 2 QSOs and liners. 
They spanned a redshift range of $0.03 < z < 2.52$ with a median of 
$\sim0.22$, therefore, the AGN were predominately local. 
The extrapolated surface density of all AGN types was $\sim0.57$ deg$^{-2}$
and is significantly higher than that of optically selected 
AGN at the same $K_s$ magnitude.
A significant fraction also showed unusually high polarization
properties \citep[][]{53} and very weak X-ray emission \citep[][]{61}.

Other 2MASS studies involved cross-correlating with the FIRST
radio catalog \citep[][]{18,17}.
These turned up a number of extremely red QSOs at high 
redshift, some showing strong evidence for dust.
\citet{1} studied the 2MASS colors of QSOs identified at 
other wavelengths, primarily from the \citet{57} catalog,
and \citet{15} used mid and far IR observations of red 2MASS
AGN to infer their relationship to Luminous IR Galaxies (LIRGs).
Most of these studies however used single band 2MASS 
detections (in $J$, $H$ or $K_s$) and therefore could have missed many red AGN.
Furthermore, none were able to provide a census of AGN 
in the 2MASS catalog. 

A uniform unbiased survey of 2MASS AGN was carried 
out by \citet{12} by selecting candidates in the
southern hemisphere covering 12.56 deg$^2$ using a moderate
color cut of $J - K_s > 1.2$. This color cut significantly reduced
contamination from foreground stars. Spectroscopic follow-up revealed that 
$\sim1.2\%$ were broad-line AGN and
$\sim4\%$ were galaxies with Seyfert 2 nuclei. 
The main findings were that: (i) the type 1 AGN are predominately 
at low redshifts ($z<~0.3$) and contamination from
host galaxy light would make them hard to find in optically selected 
QSO samples (e.g., the SDSS); (ii) the incidence of type 2 AGN
amongst local galaxies is higher than usual when compared to
blue-selected galaxy samples.
The results of this study are complementary to
those presented here and will be discussed in more detail later.

The above studies indicate that 2MASS is sensitive to the local AGN 
population. The luminosity of broad-lined QSOs in particular
in the local universe
is only slightly greater than that of their host
galaxies \citep[e.g.,][]{20}.
Only a small amount of dust extinction is needed to
mis-identify them or
turn them into type 2 AGN. This is
in contrast to the heavily obscured, high-luminosity, and usually
unresolved QSOs
at high redshift. Therefore, 
even though near-IR color-selected surveys are less biased than 
those in the optical/UV, they are still somewhat 
biased against the reddest QSOs.
They do however have the advantage that candidates selected in the near-IR will
be bright enough to allow spectroscopic followup with relative ease.

It is worth mentioning why a red $J - K_s$ color cut
has proven so efficient at finding AGN in the relatively shallow 2MASS
Point Source Catalog (PSC). There are three reasons. 
First, contamination from halo giant and disk dwarf stars is very much 
eliminated, where the majority have
$J - K_s\,\lesssim\,0.75$ \citep[][]{12}.
Second, the 2MASS colors of galaxies are predominately blue, with a median  
$J - K_s\sim1.1$ at $z\lesssim0.1$ and a 1-$\sigma$ upper limit 
of $J - K_s < 1.6$ at $z\simeq0.2$ \citep[][]{27,28}.
This color reddens rapidly for galaxies
at higher redshifts due to the $k$-correction,
approaching $J - K_s\simeq2$ at $z\sim0.4$. 
However, typical $L^{*}$ galaxies at such redshifts will be well below the 
detection limit of the 2MASS PSC. Contamination from non-active galaxies 
is therefore expected to be small overall.

The third reason is
motivated by the fact that an extremal color cut of $J - K_s > 2$ has been
shown to discriminate
red AGN from UV/optically-selected ones in
large samples of known AGN \citep[][]{1,7}.
For example, 2MASS detects a large fraction ($\sim75\%$) of the LBQS QSOs in all
three near-IR bands ($J$, $H$ or $K_s$). Most of these have $J - K_s < 2$.
Therefore, $J - K_s > 2$ isolates the reddest subset of
the optically selected population and is likely to probe many 
more.
It is also interesting to note that nearly all known QSOs at 
$z\lesssim0.5$ have $J - K_s > 1.2$ \citep[][]{11,1}.

In this paper, we extend the work of \citet{7}
and \citet{12} to search for additional red 2MASS AGN in the 
southern equatorial sky. We assembled a relatively unbiased sample
of red AGN candidates using a color cut of $J - K_s > 2$ on the 2MASS 
Point Source Working Database and then
used the brute-force capabilities of the Six Degree Field (6dF)
multiobject spectrograph to obtain spectra of a subsample.
We utlized the efficient mapping strategy of the 6dF Galaxy
Survey (6dFGS) with our candidates selected as secondary
targets in the program.

Our sample and target selection are described in Section 2.
Observations and data reduction are described in Section 3, and 
spectral classifications in Section 4.  
Properties of the newly discovered AGN and comparisons to
optically selected QSO samples are discussed in
Section 5. Conclusions are given in Section 6.
We assume a concordance cosmology with
$H_0=70\,\kms$Mpc$^{-1}$, $\Omega_m=0.3$, and $\Omega_{\Lambda}=0.7$. 
All magnitudes, unless otherwise specified, are based on the Vega system.

\section{Sample and Target Selection}

Candidates were initially selected from the 2MASS Point Source Working
Database using the following criteria: a color cut of $J - K_s > 2$;
$K_s\leq15.5$; detections in all three bands ($J$, $H$ and $K_s$);
galactic latitude $|b| > 30^\circ$; and excluding a region of
$\sim170$ deg$^2$ covering the Large and Small Magellanic Clouds.
Previously identified sources were not omitted. This yielded 16,977
candidates in an effective area of $\sim20,400$ deg$^2$ over the whole sky.
These criteria define the ``master'' catalog of red AGN candidates, and
were used in northern hemisphere follow-up studies by \citet{7}.

Note that a detection in the H-band was included for reliability. The red AGN
candidate selection criteria were originally devised during the early
stages of the survey before many of the source quality metrics were
mature. Sources detected in all three survey bands were known to be
of the highest reliability.  Therefore, three-band detection was
included with the two-band color limit to minimize sample    
contamination by spurious sources with unusual colors.

Southern equatorial ($\delta < 0^\circ$) AGN candidates from the master
catalog were then position matched against the SuperCOSMOS Sky Survey
database \citep[][]{19a}. A match radius of $4$ arcsec was used and
no optical magnitude cut was initially imposed.
$\simeq0.7\%$ of matches resulted in multiple 
(ambiguous) associations and were excluded. This yielded 6386 matches with
an overall position-difference dispersion of $\sigma\simeq0.6$ arcsec.
A sample of 2260 candidates was then compiled by selecting sources
with SuperCOSMOS magnitudes of $b_J\leq18$ and $r_F\leq17$. These 
limits were required to obtain good signal-to-noise ratio spectra
($S/N\,\gtrsim$ 10/pixel) and enable reliable identifications.
The reliability of the SuperCOSMOS optical detections 
to these magnitude limits is expected to be $>99.9\%$ \citep[][]{19b}.
The single $b_J$ and $r_F$ band photometry has an accuracy 
of $\sigma\sim0.3$ mag while due to specifics of the calibration procedure, 
$b_J-r_F$ colors are expected to have an
accuracy of $\sigma\lesssim0.12$ mag \citep[for details, see][]{19b}.
The 2260 candidates were then
proposed for follow-up with the 6dF multifibre spectrograph.
1182 were eventually allocated fibers, mainly as secondary targets during
scheduling of observations for the 6dFGS \citep[][]{29}.
Our objects were distributed over an effective
non-contiguous area of $\sim1592$ deg$^2$.

It's important to note that our initial sample of candidates
(with $J - K_s > 2$) was selected from an early version of the 
2MASS Point Source Working Database.
Subsequent recalibration and selection of alternate observations
of some of these sources for inclusion in the final
2MASS PSC resulted in some of them having colors $J - K_s < 2$.
In the end, the
majority of sources classified as AGN had $J - K_s\,\gtrsim\,1.5$,
with $\sim20\%$ satisfying 
$1.5\leq J - K_s\leq2$ according to photometry in the public-release PSC.
Uncertainties in the $J - K_s$ colors were typically 
$\lesssim0.16$ mag (1-$\sigma$).

\section{Observations and Reduction}

Spectra were obtained over the course of the 6dFGS during 2001-2006
using the UK Schmidt Telescope and the 6dF spectrograph \citep[][]{59,46}.
For details on the 6dFGS observing strategy,
see \citet{29,30}. The primary sample for the 6dFGS was drawn from
the 2MASS Extended Source Catalog \citep[XSC;][]{27}.
Seventeen additional (secondary) extragalactic samples were merged with
the primary sample \citep[see Table 3 in][]{30}. During survey
design, these were given priority indices and our initial sample of
2260 candidates had a completeness in coverage of 91.7\%.
This gave $\sim6$ AGN candidates per 6dF field
on average, although not all 6dFGS fields contained our targets because of
the different galactic latitude constraints.

The 6dF multifiber spectrograph was able to record up to 120 simultaneous
spectra over a $5.7^\circ$ field. Each fiber has a projected diameter of
$6.7$ arcsec on the sky. The 2MASS positions were accurate to
$\leq0.5$ arcsec and therefore light losses due to fiber positioning
errors were expected to be small.
For the 10$^{th}$--90$^{th}$ percentile range in redshift for the
extragalactic identifications, $z\simeq0.15$--0.45, the fiber
diameter corresponds to physical scales of $R\,\simeq\,17.5$ to
38.7 kpc $h^{-1}_{70}$. This means the 6dF spectra sampled light from
entire galaxies, and not necessarily their nuclei.

Each spectrum was taken using V and R gratings, whose outputs were later
spliced to cover the effective wavelength range:
$\sim$ 3900--7500\AA. The observed $S/N$ was typically 3-10 pixel$^{-1}$,
with $>10$ pixel$^{-1}$ being nominal given the
brightness of our sources. Spectra with low $S/N$ were primarily due to poor
observing conditions. The spectral resolution was typically
$R\sim1000$ throughout,
corresponding to emission-line
Full Width at Half Maxima (FWHM)
of $\sim$ 4--8\AA~ over the observed wavelength range. This enabled us to
resolve rest-frame velocities of $\gtrsim\,205\,\kms$ over the range
$z\simeq0.15$--0.45, sufficient for AGN identification
and classification.

\begin{figure}[h]
\begin{center}
\includegraphics[scale=1.0]{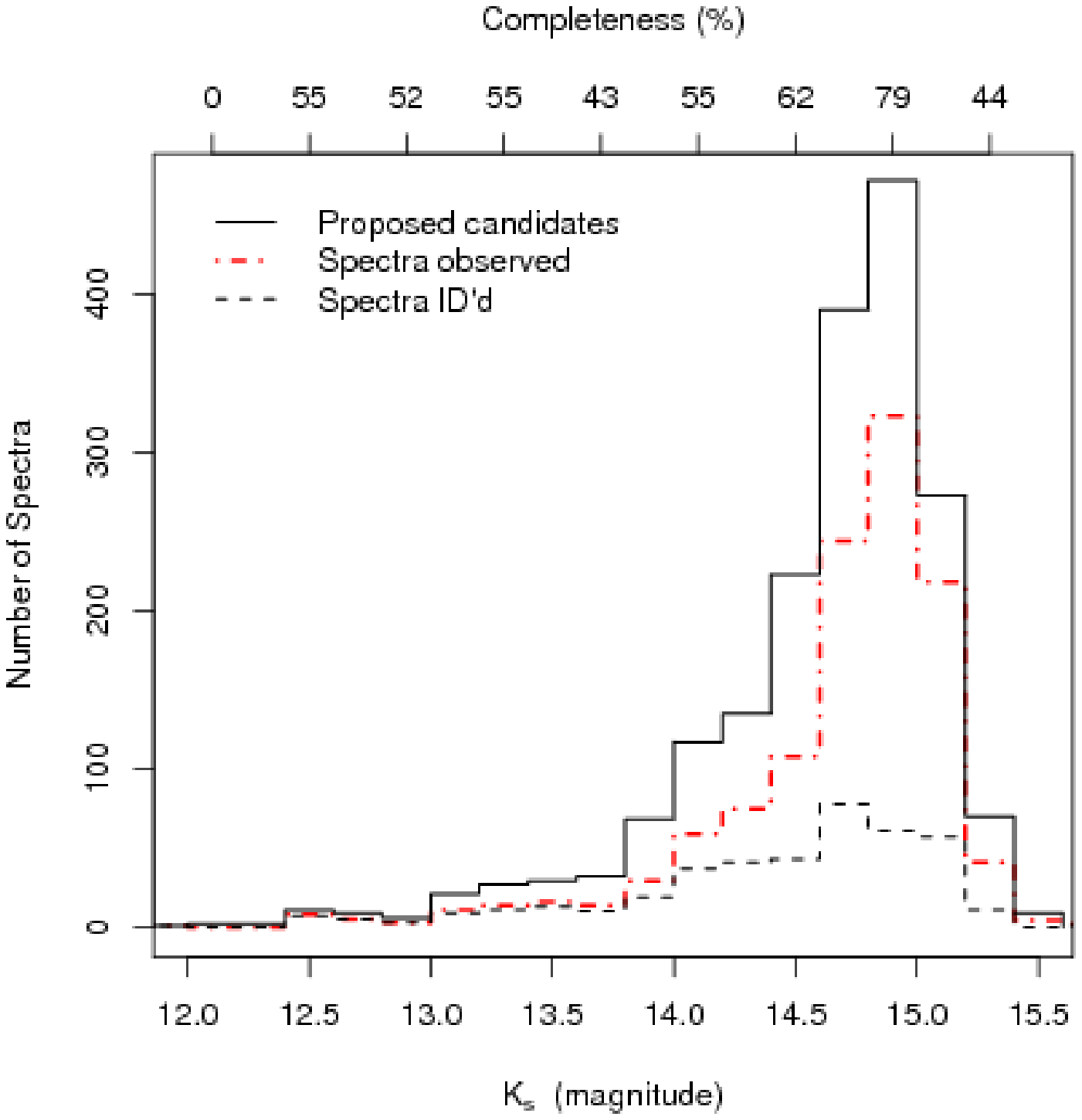}
\caption{$K_s$ brightess distribution of proposed AGN candidates,
sources with observed spectra, and sources for which we secured a
reliable spectral identification. The top horizontal axis shows the
approximate completeness ($=$ number spectra observed/number candidates)
for several magnitude bins.}\label{classHistoK}
\end{center}
\end{figure}

\placefigure{classHistoK}

The data were reduced, spectra extracted, and wavelength calibrated
using a modified version of the 2dFDR package developed for the 2dF
Galaxy Redshift Survey. Details of the reduction 
are described in \citet{29} and products
from the final data release (DR3; April 2009) are described in \citet{30}.
The flux calibration was very crude in that the same average
spectral transfer function (derived once using a couple of standards)
was assumed for every 6dF observation for all time. The spectra are
therefore not of spectrophotometric quality. This severely limited
our classification process, e.g., using emission line ratios (Section 4).
The spectra were corrected for atmospheric absorption and emission features.
However in some cases, imperfect sky-subtraction has left the imprint 
of the brightest sky lines.

Quality flags were assigned by the semi-automated 6dFGS redshift
determination pipeline (see Section 4 for details). Almost all redshifts
were visually inspected by the 6dFGS team.
Quality flags in the range $Q=$1--4
were assigned in the final public catalog\footnote{accessed via
http://www.aao.gov.au/6dFGS/}. $Q = 4$ represents a
very reliable redshift where typically the median
spectral $S/N$ was $\sim10$ pixel$^{-1}$. $Q = 3$ was assigned to
``likely'' redshift and $Q = 2$ to
tentative redshift with spectra warranting further examination.
We visually examined all spectra with
quality flags $Q\geq2$, although the majority
of usuable spectra had $Q=4$ and a handful had $Q=3$. Due to the
faintness of the sources in general, 750 of the 1182
spectra observed were of such poor quality that no classification
was possible. Classifications were therefore secured for 432 spectra.

\begin{figure}[h]
\begin{center}
\includegraphics[scale=1.0]{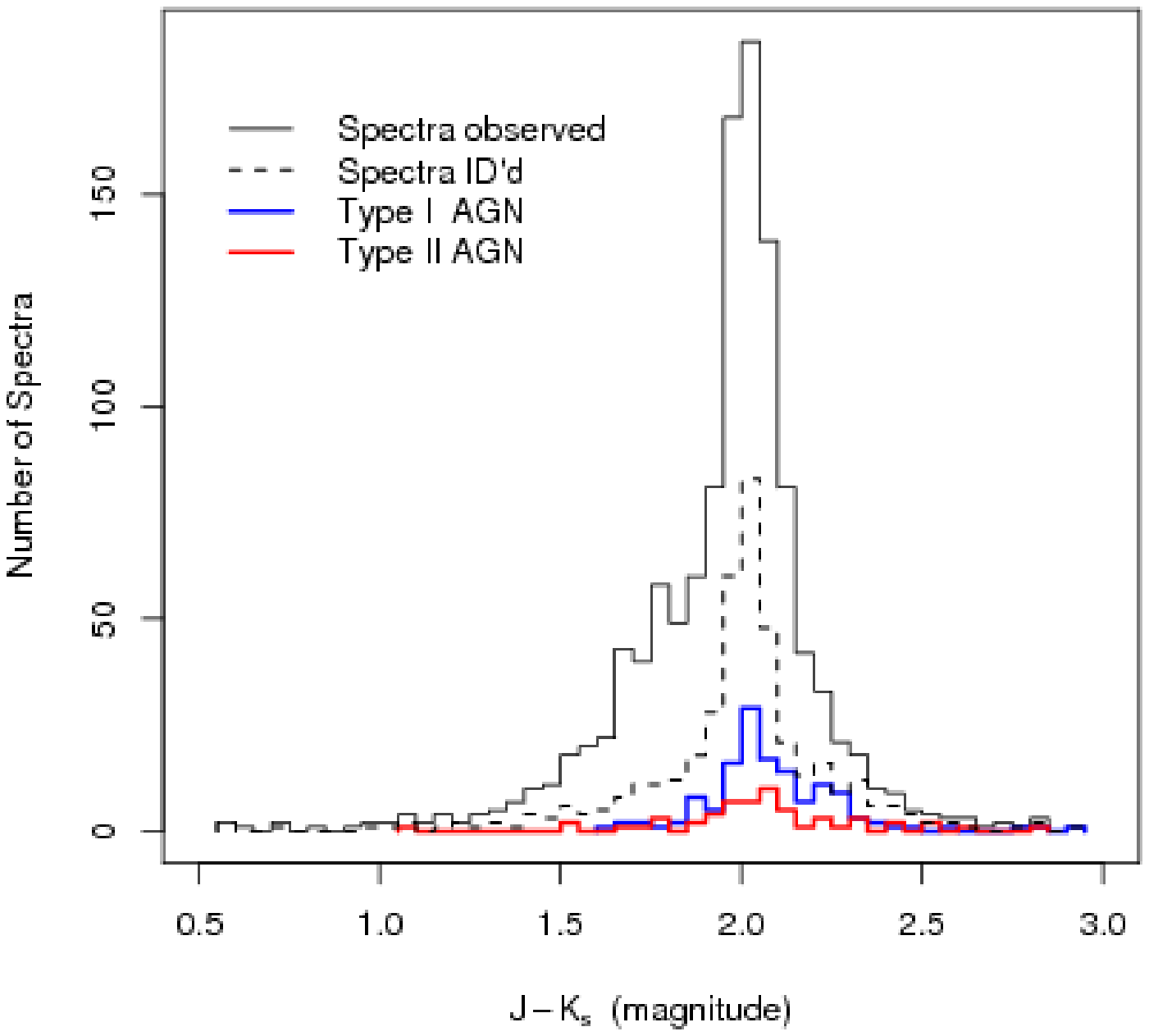}
\caption{$J - K_s$ color distribution of sources with observed spectra,
sources for which we secured a reliable spectral identification, and those
identified as Type-1 and Type-2 AGN.}\label{classHistoJmK}
\end{center}
\end{figure}

Figure~\ref{classHistoK} shows the number of proposed AGN
candidates (using the optical/near-IR constraints defined in Section 2),
the number of 6dF spectra observed, and the number with secure spectral
identifications as a function of $K_s$ magnitude.
The dearth of candidates in the faintest bin, $15 < K_s < 15.5$
is due to a combination of our optical magnitude limits (see below) and the
original $J - K_s > 2$ cutoff.
This cutoff implies
$J\,\gtrsim\,17$ for $K_s > 15$ and hence a fraction
of sources are expected to fall below the
$J$-band flux limit and excluded from the candidate list.
This drop was also seen in the \citet{7} sample of
704 candidates {\it with} follow-up optical spectroscopy.
Figure~\ref{classHistoJmK} shows that we are not
significantly biased against identifying sources with the
reddest $J-K_s$ colors. In fact, the spectral identification rate 
as a function of $J-K_s$ is approximately uniform.

\begin{figure}[h]
\begin{center}
\includegraphics[scale=1.0]{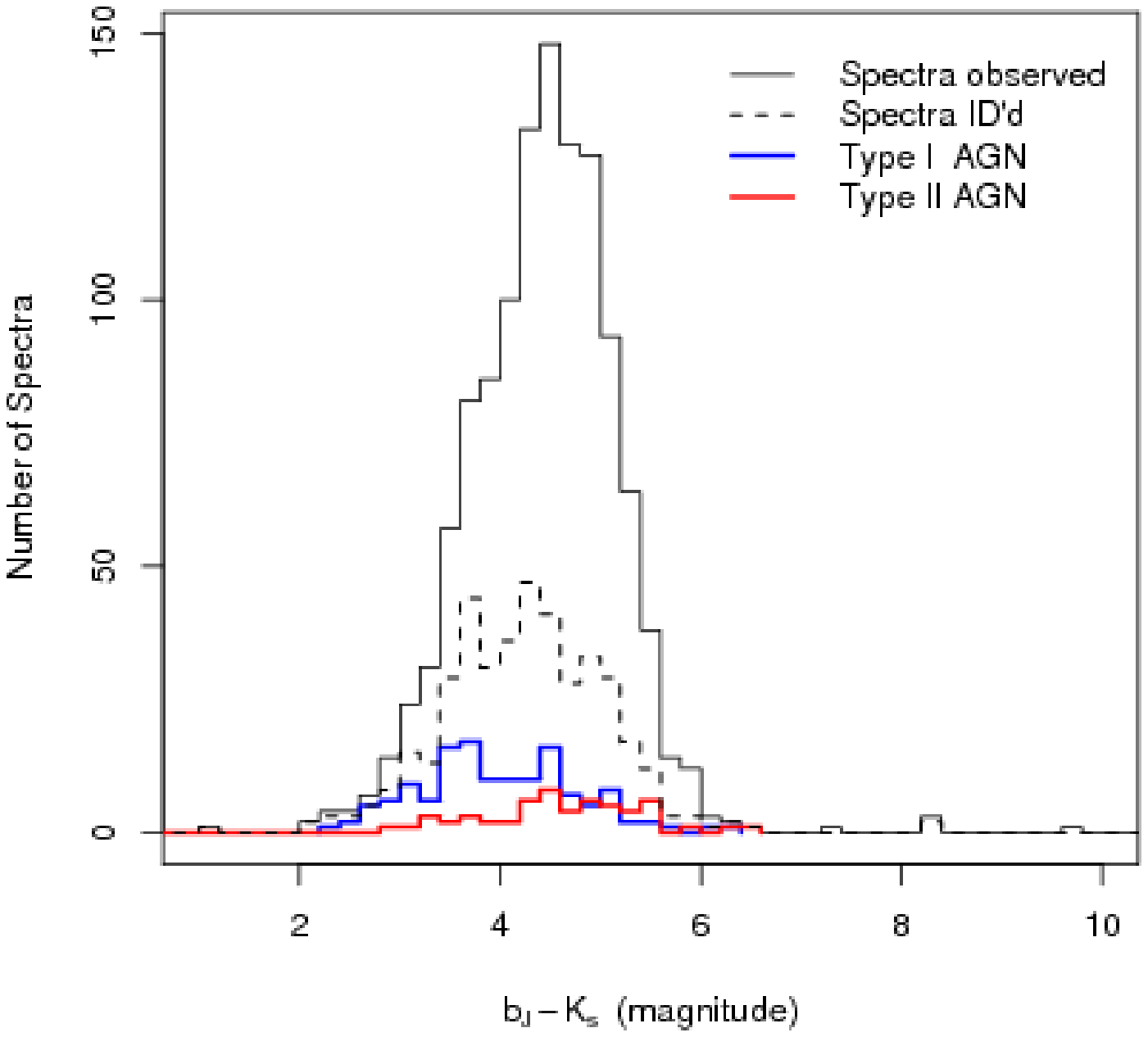}
\caption{$b_J - K_s$ color distribution of sources with observed spectra,
sources for which we secured a reliable spectral identification, and those
identified as Type-1 and Type-2 AGN.}\label{classHistoBmK}
\end{center}
\end{figure}

Figure~\ref{classHistoK} shows that
the number of spectra observed uniformally samples the
$K_s$ distribution of candidates, i.e.,
the completeness is approximately uniform. However, there is relatively
higher incompleteness in the number of spectra {\it identified}
at the faintest magnitudes: $14.5 < K_s < 15.25$. The incompleteness
in this range is $\sim70\%$ with respect to the number of spectra observed.
This is primarily due
to the faintest sources generally having poorer quality spectra.
These are expected to be near the optical magnitude limit imposed for
spectroscopy: $17\,\lesssim\,b_J\,\leq\,18$.
In fact, the introduction of an optical magnitude cut
is expected to have biased the spectral sample towards
bluer optical colors in general. 
Comparing the relative deficit in the number of faint ($K_s > 14.5$) 
sources to the original 
candidate $K_s$ distribution \citep[from][]{7} with no 
optical magnitude limit imposed, we estimate we have lost $\gtrsim\,35\%$ 
of candidates due to this optical cut. A majority of these missed
candidates (with $b_J>18$) have $b_J-K_s$ colors $\simeq4-6$.
However, the {\it spectrally}-observed $b_J-K_s$ distribution
(Figure~\ref{classHistoBmK}) shows that we are not completely 
biased against identifying the reddest sources.
When compared to the colors of optically selected QSOs, 
the 2MASS AGN have a tail extending to moderately redder colors
(see Figure~\ref{classJKvsBK} and Section 5.2 for more details).

\placefigure{classHistoJmK}

\placefigure{classHistoBmK}

\section{Classification}

All spectra were initially classified using the semi-automated
6dFGS classification software, whose primary purpose was to determine
accurate redshifts. This is a modified version of the {\scriptsize RUNZ}
software used for the 2dF Galaxy Redshift Survey \citep[][]{5}.
It used 13 spectral templates to identify spectra using line-fitting
and cross-correlation techniques. This program produced very reliable
redshifts for all galaxies in general, although was less reliable at
separating out the various AGN classes from late and early type
galaxies, and stars. All spectra were visually inspected to determine
whether a poor-quality spectrum flagged by the 6dFGS software was worthy
of further examination.

432 spectra were of sufficient quality
to enable a classification of some sort, but not necessarily an
unambiguous identification.
All spectra were shifted to their rest-frame using redshifts
determined by the 6dFGS program.

\subsection{Emission-Line Diagnostics}

We assembled a database of emission-line diagnostics for all the good
quality spectra to assist with the identifications. The diagnostics included
line fluxes, equivalent widths (EWs), and dispersion velocities.
These were estimated by fitting Voigt profiles and the underlying
continua were approximated by linearly interpolating straight-line fits 
on either side of each line. Line fluxes were then determined by 
integrating the flux in the fitted profiles above the continuum level.
Dispersion velocities were determined from the FWHM of the lines.
The lines of interest and the effective
wavelength regions used to define the continuum and line integration limits
are shown in Table~\ref{lines}.

\placetable{lines}

The {\it fitprofs} task in {\scriptsize IRAF} was used for the automated
measurement of line diagnostics. This included the ability to deblend
closely separated lines, e.g., H$\alpha$ and [N {\scriptsize II}].
An important input parameter for the {\it fitprofs} task is an estimate of
the 1-sigma uncertainty per pixel. This allows 
the program to compute uncertainties in each of the fitted
line parameters. Since the fitting was non-linear, this was accomplished
using a Monte Carlo simulation.
The spectral uncertainty was computed by first selecting
a relatively clean region in the rest frame common to each spectrum,
i.e., devoid of strong emission and absorption lines. We selected the
rest wavelength range 5050--5400\AA.
We then fitted a straight line to the data in this region
using a robust (outlier-resistant) regression method based on
the general concept of ``M-estimation'' \citep[][]{25}. The uncertainty
was then estimated using the median absolute deviation in the residuals
from the fit, and is also robust against potential outliers:
\begin{equation}
\sigma\simeq1.4826\,median\left\{|p_i - fit\{p_i\}|\right\},
\end{equation}
where $p_i$ is the value of the $i^{th}$ pixel in the 1-D spectrum
and $fit\{p_i\}$ is the fitted value. This quantity is scaled such that it
converges to the standard-deviation of a Gaussian in the limit of a large
sample. For spectra where the range 5050--5400\AA~ included or fell outside
the long wavelength end ($\lambda_{red}$) after shifting to the rest frame,
a wavelength range of $\lambda_{red}-250\leq\lambda\leq\lambda_{red}-20$\AA~
was used instead. In all cases, this ensured $>20$ pixels for the noise
computation.

\begin{figure}[h]
\begin{center}
\includegraphics[scale=1.0]{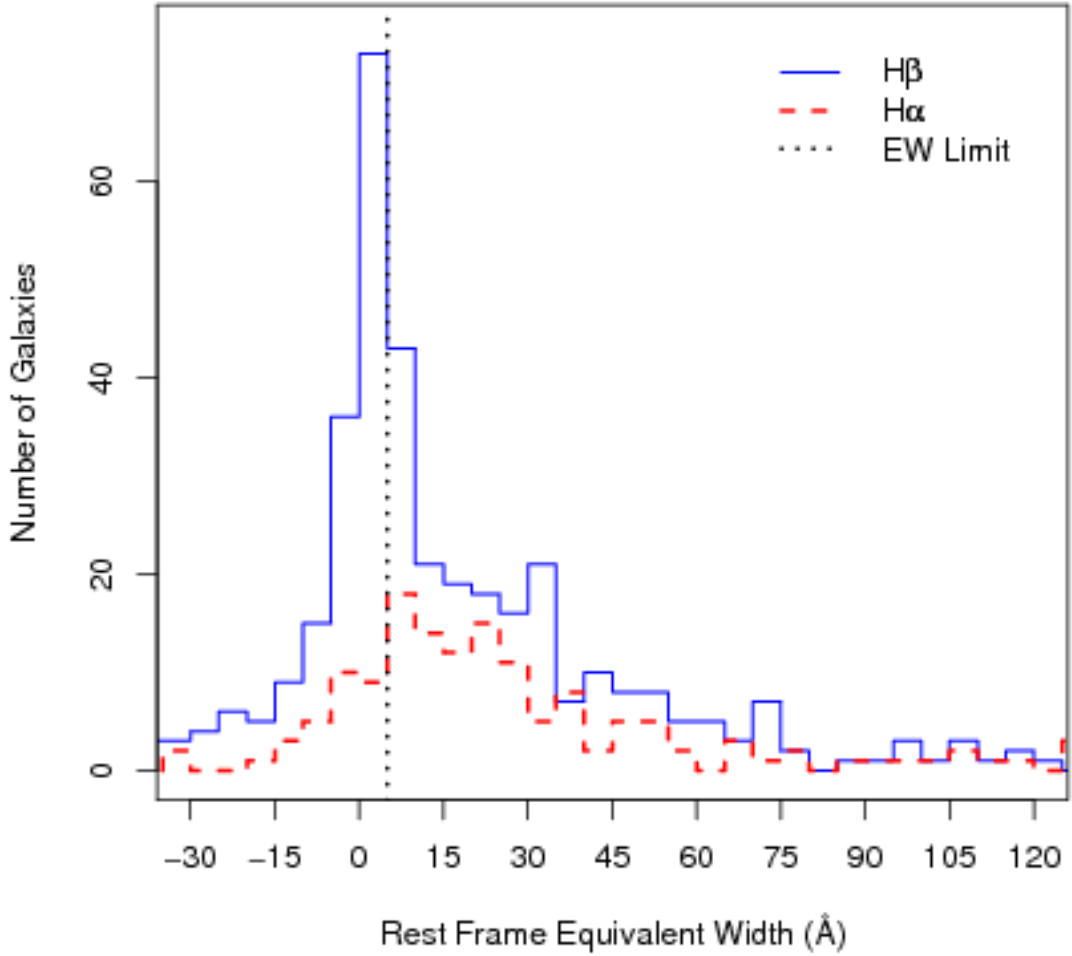}
\caption{Distribution of rest-frame H$\alpha$ and H$\beta$
equivalent widths for all galaxies with the best quality
specta. We are sensitive to all equivalent widths to the right
of the vertical dotted line ($>5$\AA).}\label{classHistoHbHa_ew}
\end{center}
\end{figure}

We estimated the smallest EW we are sensitive to
by examining the dispersion in H$\alpha$ and H$\beta$ EWs 
of all the galaxies (and potential AGN) with the best quality 
spectra ($Q = 4$). Figure~\ref{classHistoHbHa_ew} shows 
these distributions.
EW measurements clustered around zero are lineless galaxies where
an identification would be highly unreliable, if at all possible.
We find we should be sensitive 
to galaxies with rest frame EW in either H$\alpha$ or 
H$\beta$ of $\gtrsim\,5$\AA. Given our observed wavelength range,
either of these lines are expected to be observed
at $z\lesssim0.5$ and therefore were used as
constraints in the identification process below.

\placefigure{classHistoHbHa_ew}

Line fluxes and EWs were also measured interactively by integrating the line
fluxes directly from the 1-D spectra. These were in excellent agreement,
to within measurement error, with the profile-derived fluxes from above.
All lines were visually inspected and unreliable flux measurements flagged.
These were primarily lines that were contaminated by a strong sky line or
atmospheric absorption band. Our final emission line database retained
lines with fluxes $\gtrsim\,2.5\sigma$, and lines that were
clearly discernable by eye in case our automated measurement
of $\sigma$ was overestimated. Unreliable $\sigma$ estimates
occured in $\simeq7\%$ of the spectra.
Note that we did not correct the emission line fluxes for any underlying
absorption (e.g., from stellar photospheres) since a study using similar
spectra from 2dF by \citet{12} found this effect to be negligible.

A first pass examination of the spectra together with initial
classifications provided by the 6dFGS program motivated us to define
six broad object classes: type 1 AGN; type 2 AGN; starburst or late-type
star forming galaxies; early type galaxies; stars; and unknown emission
line galaxies.  These classes and the criteria
used to identify them are as follows.

Spectra with broad H$\alpha$ and/or H$\beta$ emission lines
exceeding 1000 $\kms$ (FWHM), or with other broad
permitted lines present, e.g., C {\scriptsize III}]
or Mg {\scriptsize II} for $z\,\gtrsim\,1$ and $z\,\gtrsim\,0.4$ respectively,
were classified as type 1 AGN.
Included in this criterion are $S/N\,\geq\,2.5$ on the FWHM measurement and a
H$\alpha$ or H$\beta$ EW $> 5$\AA. Spectra in
which known broad lines could be discerned by eye but were relatively noisy,
i.e., with flux $S/N < 2.5$ were classfied as ``probable'' type 1 AGN.

\begin{figure}[h]
\begin{center}
\includegraphics[scale=1.0]{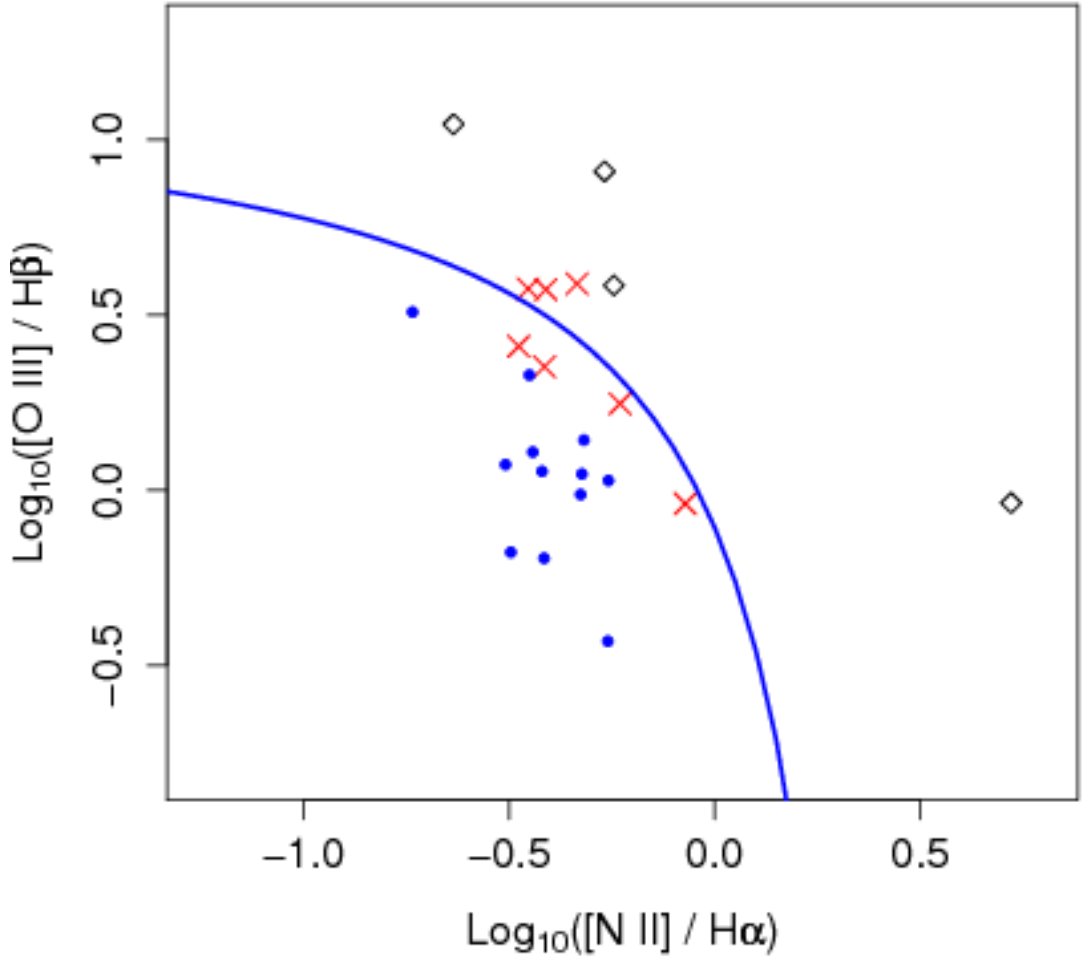}
\caption{Line ratio diagram for galaxies with type 1 
AGN classifications removed.
Diamonds and filled
circles are ``probable'' type 2 AGN and star-forming galaxies
respectively, all at $\geq0.2$ dex (1-$\sigma$) from the
classification boundary of \citet[solid line]{32}.
Crosses are composites (classified as unknown galaxies).
Errors in the line ratios are
typically 0.2 dex (1-$\sigma$).}\label{classNIIflxratio}
\end{center}
\end{figure}

\begin{figure}[h]
\begin{center}
\includegraphics[scale=1.0]{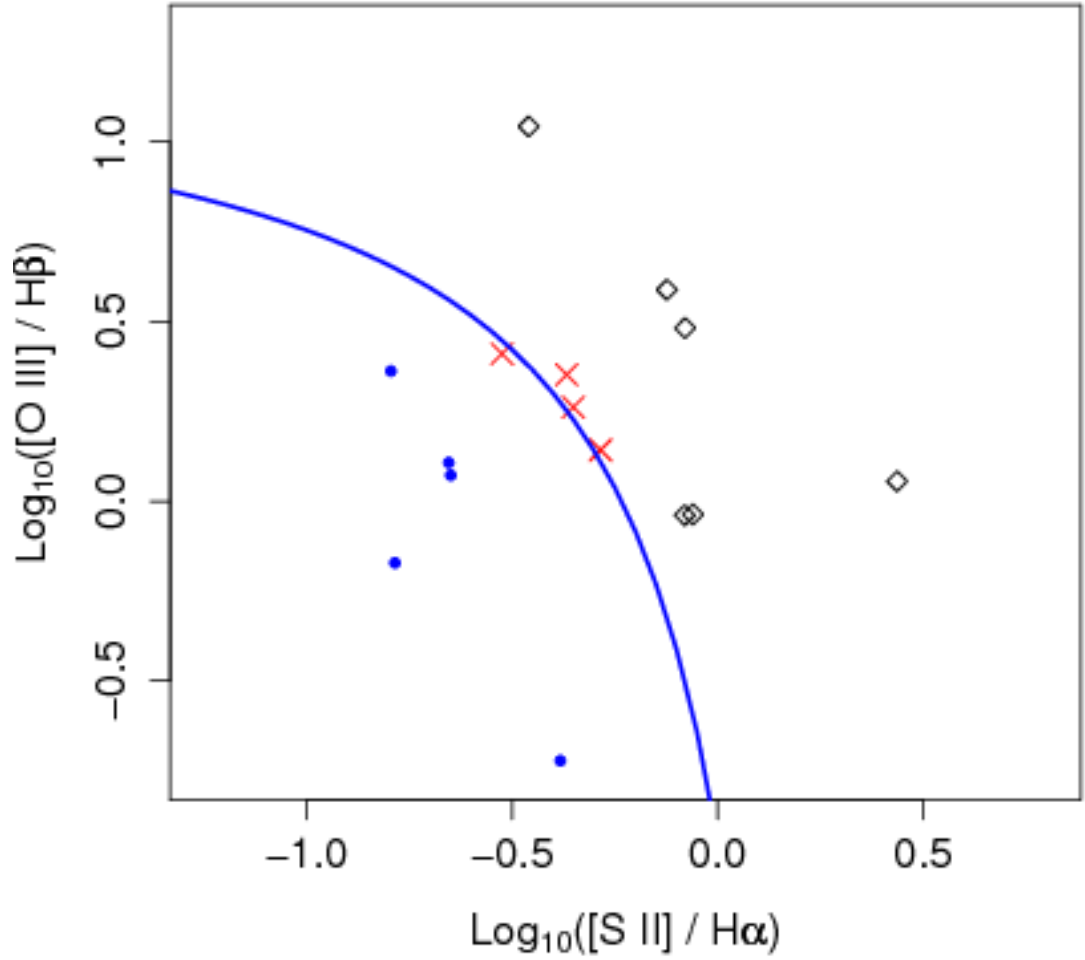}
\caption{Same as Figure~\ref{classNIIflxratio} but with
[S{\scriptsize II}] in the abscissa. The solid line is
the AGN/star-formation discriminator from \citet{32}.
Errors in the line ratios are
typically 0.2 dex (1-$\sigma$).}\label{classSIIflxratio}
\end{center}
\end{figure}

Non-type 1 AGN spectra were classified using line ratios involving
good measurements in either of the following line pairs:
([O {\scriptsize III}], H$\beta$) or ([N{\scriptsize II}], H$\alpha$) or
([S{\scriptsize II}], H$\alpha$) or
([O {\scriptsize III}], [O {\scriptsize II}]).
We first attempted a classification
using the diagnostic diagrams of \citet{32,33}, which are based
on the classic BPT diagrams of \citet{33a}.
Figures~\ref{classNIIflxratio} and~\ref{classSIIflxratio} show the
traditional line-ratio diagrams using our good quality spectra
(with pre-classified
type 1 AGN removed) and where {\it all} four lines had flux $S/N\geq2.5$.
Unfortunately, all four lines in either Figure~\ref{classNIIflxratio}
or ~\ref{classSIIflxratio} were only simultaneously visible
(and with good $S/N$) in
$\simeq\,$8\% (23/296) of the good-quality non-type 1 spectra.
Furthermore, the errors in the line ratios
($\simeq\,$0.2 dex, 1-$\sigma$) were too large
for the bulk of these spectra to be reliably classified.
We therefore declared the few type 2 AGN and star-forming galaxies
that could be classified using this method
(at distances $\geq$ 1-$\sigma$ from the classification
boundaries) to be probable identifications. 

\placefigure{classNIIflxratio}

\placefigure{classSIIflxratio}

For the remaining 273 sources with good quality data and where
only one of the above line pairs was available (predominately when
$z\,\gtrsim\,0.1$),
a galaxy was classified as a ``probable''
type 2 AGN if either of the following was
satisfied: log([O {\scriptsize III}]/H$\beta)>0.3$ or
log([N {\scriptsize II}]/H$\alpha)>-0.2$ or
log([S {\scriptsize II}]/H$\alpha)>-0.35$ \citep[e.g.,][]{62} or
log([O {\scriptsize III}]/[O {\scriptsize II}]) $>0$
\citep[e.g.,][]{13,33}.
Combined with any of these, we also required a
rest frame FWHM([O {\scriptsize III}]) $>300\,\kms$,
FWHM(H$\alpha$ or H$\beta)<1000$ $\kms$, and
H$\alpha$ or H$\beta$ EW $>$ 5\AA.
The FWHM([O {\scriptsize III}]) limit was included to improve the  
relability of type 2 identifications when only
one line pair was available. For comparison,
\citet{62} assumed FWHM([O {\scriptsize III}]) $>400\,\kms$.
We assumed $300\,\kms$ since the
distribution for FWHM([O {\scriptsize III}])
for their entire type 2 sample
falls off sharply at $<300\,\kms$. In the end, this limit made little 
difference to the type 2 identification statistics.

Probable starburst/late-type starforming galaxies were classified using the
negation of these single line ratios with some buffer to allow
for flux errors, i.e., with:
log([O {\scriptsize III}]/H$\beta)<0.2$ or
log([N {\scriptsize II}]/H$\alpha)<-0.3$ or
log([S {\scriptsize II}]/H$\alpha)<-0.4$ or
log([O {\scriptsize III}]/[O {\scriptsize II}]) $<-0.2$.

\placetable{summ}

It's important to note that the type-2 AGN identified using the
above single line
pairs could be contaminated by low-metallicity
emission-line galaxies or liners,
especially at low redshift. Furthermore, the non-spectrophotometric
nature of our spectra could invalidate some identifications made using
widely separated lines (e.g., the pair [O {\scriptsize III}],
[O {\scriptsize II}]). These classifications are therefore very tentative
given the quality of our data. We therefore declare all type 2 AGN
identifications quoted in this paper to be probable. Follow-up with
higher $S/N$ spectral observations, preferably with better calibrated
throughput as a function of wavelength will be needed for confirmation.

Early type galaxies were identified through the characteristic
4000 \AA~ break, a signature caused by the dearth of hot and young
(usually O and B-type) stars and strong heavy metal absorption by
stellar photospheres.
Stars were isolated by first ensuring that their radial
velocities were $\lesssim150\,\kms$. Their spectra were then matched to
templates from the ELODIE stellar library \citep[][]{41}.
The majority were K and M dwarf stars. All other galaxy-like spectra
with H$\alpha$ or H$\beta$ EW $>$ 5\AA~ but not fitting the above criteria,
e.g., with single line measurements, or composites lying within 1-$\sigma$
($\pm0.2$ dex) of the type 2/starburst classification boundaries in
Figures 5 or 6 were classified as ``unknown galaxies''.

\subsection{Results Summary}

Table~\ref{summ} summarizes our source classifications.
We only include our secure identifications for the type 1 AGN with
probable identifications (as described in Section 3)
placed in the ``Unknown Galaxies'' class.
Also included are
statistics for type 1 and type 2 AGN from the \citet{7}
northern 2MASS AGN survey, and the \citet{12} southern
2MASS AGN survey. The latter is broken into two color cuts.
The \citet{7} study has at least twice the
detection rate for type 1 AGN. This could be due to deeper spectroscopic
follow-up of fainter single targets in their study. It is possible that
a significant fraction of our fainter targets at $K_s>14.5$
(e.g., Figure~\ref{classHistoK}) where a spectral identification
could not be secured are type 1 AGN.
It's also interesting to note that
there is a tendency of the type 1 AGN identification rate to increase with
$J-K_s$ color in Table~\ref{summ}.

Figure~\ref{spectra} shows a sampling of spectra for the new type 1 AGN,
primarily those with the highest redshifts. Two of the sources are at
$z > 1$: 2MASS J21571362-4201497 with $z=1.321$ and 2MASS J10012986-0338334
with $z=1.389$. The latter has been classified as a probable type 1 AGN due to
a low spectral $S/N$, although given its relatively high redshift,
it is most likely a QSO.

Table~\ref{type1agn} lists the secure (T1) and probable (PT1) type 1 AGN
identifications. There are 116 classified as T1 and 20 as PT1.
Previous or alternative names as listed in the NASA Extragalactic
Database (NED) are given. Of the 136 type 1 AGN, 8 (or $\sim$6\%)
were previously classified as either ``AGN'', ``QSO'' or ``AGN/QSO?'' in NED.
Four of these are in the SDSS QSO sample.
This implies a majority are new, previously undiscovered AGN.
Interestingly, 10 of our type 1 AGN were previously
detected in X-rays by the {\it ROSAT} All-Sky Survey (RASS).
Table~\ref{type2agn} lists the type 2 AGN, all of which are classified as
probable using the methods described in Section 3. 
None of the type 2s were previously classified as AGN-like, and only one
was detected in X-rays by the RASS.
Of the previous classifications available in NED,
a majority of our type 1 and 2 AGN
are listed as galaxy-like and extended in the 
optical (SuperCOSMOS digitized plates) or near-IR (2MASS Atlas Images).
At least $30\%$ are also in
the 2MASS Extended Source Catalog \citep[XSC;][]{27}.
This is expected given the depth of our sample.

\placefigure{spectra}

\placetable{type1agn}

\placetable{type2agn}

\placefigure{classHistoZ}

\section{Discussion}

This section reviews the properties of our 2MASS AGN and compares them 
to those of AGN/QSOs discovered in optical surveys.
We explore their redshift, luminosity, photometric and
line equivalent-width distributions.
Our primary benchmark and comparison sample of 
optically-selected AGN/QSOs is the SDSS Quasar Catalog Data Release 5 
\citep[DR5;][]{47}.
This catalog contains 2MASS matches to 9824 AGN, all within 2 arcsec.

\subsection{Redshift and Luminosity Distributions}

Our 2MASS AGN span the range $0.01\lesssim z\lesssim1.38$ as shown in
Figure~\ref{classHistoZ} ({\it left}). The median $z$ is $\simeq0.27$ and
$\simeq0.21$ for type 1 and type 2 AGN respectively.
A majority of our AGN are at low redshifts, with only two at $z > 0.7$.
There are six securely identified AGN (4 type 1's, 2 type 2's) at $z < 0.05$.
This is 24\% (6/25) of all secure {\it galaxy}-like spectral identifications 
(including unknown types) in this redshift range.
This is a lower limit since some objects classified as ``unknown'' could
be type 2 AGN. Even removing our ``probable'' type 2 AGN, the AGN fraction
is still relatively high.
For comparison, \citet{20}
find that $\approx4\%$ of $\simeq$15,200 SDSS-detected 
galaxies at $z < 0.05$ harbor
AGN (mostly Seyferts).
An earlier study by \citet{26} found AGN (including LINERs) in
$\simeq3.4\%$ of a sample of 2399 nearby blue-selected galaxies. 
These studies are not a fair
comparison since the galaxies 
were optically selected. It would be of
interest to determine
the fraction of SDSS galaxies with active nuclei 
for a color cut of $J - K_s > 2$.

We have a significantly higher AGN identification rate than any
previous low redshift galaxy sample. This is expected to be due to
our red $J - K_s$ color selection. The \citet{12} survey
of 2MASS AGN had a bluer color cut, $J - K_s > 1.2$, and had a
significantly lower AGN fraction
(see Section 4.2 and Table~\ref{summ}).
Evidence for (low-luminosity) AGN activity was recently found in $17\%$ of a
sample of 64 late-type spiral galaxies
by \citet{8} using X-ray data from {\it Chandra}.
\citet{61} and \citet{35} 
also showed that the 2MASS red AGN are generally 
weak X-ray emitters, with the reddest $J - K_s$ sources being the weakest.
AGN down to low luminosities are therefore more
common than previously thought.

Another consideration is that the 6dF fiber
diameter corresponds to sampling physical scales of $R>12.3$ kpc $h^{-1}_{70}$
at $z\geq0.1$. This means the 6dF spectra are sampling light from 
entire galaxies, even at the lowest redshifts.
Furthermore, we are only sensitive to AGN with H$\alpha$ or H$\beta$
EW $>$ 5\AA. \citet{23}
showed that we are likely to miss many 
AGN at this EW limit within our large aperture.
From spectral observations of the nuclear regions ($\lesssim200$ pc)
of a large sample of blue-selected nearby galaxies, they found that 
almost 50\% contain active nuclei.
Even though galaxy light can significantly dilute the contribution from an AGN,
we find that a near-IR selected sample with a red color cut can reduce the 
fraction of objects whose SEDs are dominated by host galaxy light,
contrary to the claim of \citet{23}.

Our redshift distribution is consistent with
that from \citet{7} who used a similar $J - K_s$ color cut. They
detected 2 QSOs with $z > 2.3$, consistent with an enhancement of the $K_s$
band flux from H$\alpha$ emission moving into that band.
Figure~\ref{classHistoZ} (right) shows $J-K_s$ versus redshift for our
AGN and SDSS QSOs.
This shows that our red near-IR color criterion biases AGN selection
towards low-redshifts because of the $k$-correction effect of the AGN/QSO
SED. Most AGN show a sharp rise in
flux in the rest-frame between 1 and 2$\mu$m, possibly from hot dust emission
\citep[e.g.,][]{45}, and this is redshifted out of the $K_s$ band
at $z\gtrsim0.5$. The $J-K_s$ colors of SDSS QSOs and the prediction for
radio-quiet QSOs from the template of \citet{10} confirm this
trend. \citet{1} showed that 2MASS has the sensitivity
to detect QSOs with $J - K_s > 1.5$ out to $z\simeq4$, maybe higher
(see their Fig. 5), although they are very rare.

Figure~\ref{classKvsZ} compares the $K_s$ flux and luminosity between
active and inactive galaxies as a function of redshift.
We assumed a power-law SED $f_{\nu}\propto\nu^{-\alpha}$
for the $k$-correction. The slope $\alpha$ was
derived from the $J - K_s$ color of each source.
No foreground redenning correction was applied
since the extinction is typically
$A_K < 0.05$ mag for galactic latitudes $|b| > 30^{\circ}$ \citep[][]{48}.

Figure~\ref{classKvsZ} (right) shows that the near-infrared luminosities
of some of our active galaxies are comparable to those of
securely identified inactive galaxies. These are
a mixture of early-type and late-type starburst galaxies and their
near-IR emission is dominated by the less-luminous host galaxy.
Our AGN have a tail extending             
to higher luminosities (by $\simeq2$ mag) than the inactive galaxies.
This trend has been found by many authors at other wavelengths
\citep[e.g.,][]{26,20}.
Also, host galaxy emission could be non-negligible
in the type 2 AGN, suggesting they would be slightly less-luminous 
than the type 1s. This luminosity dependence is consistent with
the observation that the fraction of type 1 AGN in our sample increases
with redshift relative to inactive galaxies and there
is a dearth of type 2 AGN at $z > 0.4$. For comparison, the bulk of
the type 1 AGN extend to $z\simeq0.6$.

\placefigure{classKvsZ}

The $K_s$ luminosities of our 2MASS AGN generally overlap
with those of SDSS QSOs in the same redshift range, 
but the bulk at $z>0.2$ are more luminous on average than
the locus formed by the SDSS QSOs (Figure~\ref{classKvsZ} - {\it right}).
Our AGN therefore have luminosities closer to QSOs than
Seyferts, although we caution that the SDSS QSOs and 2MASS
AGN were selected using entirely different techniques.
Figure~\ref{classKvsZ} ({\it left}) shows that
our AGN have a $K_s$ flux limit ($K_s = 15.5$) brighter by
$\gtrsim\,0.5$ mag than the bulk of SDSS QSOs
at similar redshifts. A large fraction of the SDSS QSOs at $z\,\gtrsim\,0.2$
are detected to fainter $K_s$ magnitudes. The lower number of 2MASS AGN at
$K_s>15$ is due to the higher incompleteness in
spectral identifications at the faintest magnitudes.
We are therefore sensitive to the most near-IR luminous
objects of the optically selected QSO population.

The ratio of type 1 to type 2 AGN in our sample is $\simeq 2:1$.
This ratio cannot be compared to studies at other wavelengths due
to the high level of incompleteness in our sample, e.g., brought
about by the relatively bright optical cut
imposed by spectroscopy (see Section 3). However, compared
to previous 2MASS AGN studies that performed follow-up 
spectroscopy to similar optical limits, there 
is tentative evidence that the type 1 to type 2 ratio depends on
$J - K_s$ color in the sense that a redder color cut has
a higher proportion of type 1s. For example, \citet{12} find ratios
of 14:23 and 4:0 for $J - K_s > 1.2$ and $>1.8$
respectively, and \citet{7} found
$\simeq 4:1$ for $J - K_s > 2$ (Table~\ref{summ}).
We find $\simeq 2:1$ for an effective $J - K_s\,\gtrsim\,1.7$.
All these studies follow the same qualitative trend and
cannot be explained by redshift or luminosity dependent biases.
This is consistent with the notion that
most type 2 AGN (e.g., Seyferts 2s) have their optical-to-near-IR emission 
dominated by ``blue'' host galaxy light, and that a red $J - K_s$ cut
will select more sources where the active nuclear emission
dominates, i.e., type 1 AGN and QSOs at moderately low redshift.

\placefigure{classJKvsBK}

\subsection{Broadband SEDs and Dust Reddening}\label{SEDs}

Figure~\ref{classJKvsBK} compares the optical-to-near-IR colors of
2MASS red AGN to those of
SDSS QSOs. The SDSS optical magnitudes were converted to equivalent UKST
photographic $b_J$, $r_F$ magnitudes by first
converting them to Cousins $B$, $V$ using the
color corrections in \citet{14}, and then to UKST magnitudes
(on the Vega system) using the corrections in \citet{3}.
Overall, the type 1 AGN span a range in $b_J - K_s$ and
$b_J-r_F$ color similar to those of
optically-selected SDSS QSOs at the same $J-K_s$ color cut.
This is not surprising since our sample required a relatively bright 
optical magnitude cut for reliable follow-up 
spectroscopy (see Section 2). We could indeed be sampling the same
AGN population detected in the optical and spanning 
the same (low) redshift range
(Figure~\ref{classHistoZ} - {\it right}).
Interestingly, the spread in $b_J - K_s$ and $b_J - r_F$ of the 2MASS red AGN
are also consistent with the radio-loud quasar selected samples of
\citet{60} and \citet{11}, although these samples
spanned a large range in redshift and very few sources were detected
in all 2MASS bands. 

The type 2 AGN have redder optical colors than the bulk spanned by type 1s
(Figure~\ref{classJKvsBK} - {\it right}) where $\Delta (b_J - r_F)\simeq1$ mag.
This is consistent with the AGN unified model
where our view to the nuclear region is completely obscured
and red host galaxy light will dominate the continuum flux at all wavelengths.
The starlight can be instrinsically red (e.g., evolved stars), reddened 
by dust, or both.
The type 1 AGN have bluer optical colours, consistent with the SDSS QSOs.
The optical properties of
2MASS red type 1 AGN are therefore not dramatically different from those in
optically selected samples.

Our spectral observations are not sufficiently 
spectrophotometric to warrant use of
emission line ratios to constrain the amount of dust reddening. However,
the distribution of optical and near-IR colors does not exclude mild
amounts of dust reddening. Figure~\ref{classJKvsBK} shows reddening
vectors for dust
with an optical depth $\tau_{\lambda}\propto 1/\lambda$ applied to
the colors of
a typical blue QSO in the rest frame ($z=0$) and as observed at the
median redshift of our type 1 AGN ($z=0.27$).
In order for a ``blue'' QSO to be promoted to the region of color space
occupied by the 2MASS red type 1 AGN, we require extinctions
of $A_V < 2$ mag.
Note that there is evidence that the color-color locus
occupied by the SDSS QSOs may already be
extended by dust reddenning with $A_V\lesssim1.5$ mag
\citep[][]{44}. This is to be compared to QSOs discovered in
earlier optical/UV surveys that used more
rigid selection criteria. For example, the LBQS \citep[][]{22} 
spans $0.6\,\lesssim\,J - K_s\,\lesssim\,1.4$,
$2\,\lesssim\,b_J - K_s\,\lesssim\,4$, and
$0.5\,\lesssim\,b_J - r_F\,\lesssim\,1.5$, considerably bluer than
the SDSS QSOs. Using the LBQS as an unreddened comparison sample would
make the 2MASS AGN colors more difficult to reconcile with
a simple screen extinction model with
$\tau_{\lambda}\propto 1/\lambda$. Most the 2MASS AGN are too blue in their
optical colors for dust to be wholly responsible for modifying
their optical/near-IR continua relative to blue-selected QSOs.

However, there is some evidence that at least some 2MASS AGN are
reddened by dust.
The SEDs for a sample of 10 very red 2MASS AGN with $J - K_s > 2$,
$R - K_s > 5$ ($b_J - K_s\gtrsim5.5$) were modeled by \citet{15}.
They found that all sources were consistent with moderate amounts of
dust redenning of $A_V = 1.3-3.2$ mag. They also found that the mid-IR SEDs
are dominated by hot dust and that their 60/12$\mu$m luminosity ratios are
significantly higher than those of blue-selected Palomar-Green
(PG) QSOs \citep[][]{49}, suggesting a higher level of star-formation.
Similar conclusions were reached by \citet{35}, including
evidence that in a handful of highly polarized objects, the optical has  
a contribution from AGN light scattered by dust.

\placefigure{classSEDs}

We repeated a similar analysis as in \citet{15} to determine
whether our 2MASS red AGN have a significantly higher far-IR emission
on average. A search through the {\it Spitzer} and {\it ISO} archives turned up
no mid-to-far IR data. However, eight type 1 AGN were securely detected by
{\it IRAS} in all bands (12, 25, 60, and 100$\mu$m).
One of these (2MASS J04225656-1854422) was previously identified as 
a Serfert 1 galaxy, another with an Ultra-Luminous IR Galaxy that was also 
detected in the RASS (2MASS J02460800-1132367), and the remainder
are given a ``Galaxy'' classification in NED.
The eight IRAS all-band-detected AGN span redshifts
$0.06\leq z\leq0.27$ and are at the red end of the $b_J - K_s$ color
distribution with $b_J - K_s\,\gtrsim\,4.5$.
Photometry was extracted from the {\it IRAS} Faint Source
Catalog V2.0 using the IPAC Infrared Science Archive. In fact,
only 10\% of our 2MASS AGN (both type 1 and 2) were detected in at least
one band by {\it IRAS}, indicating the 2MASS red AGN are not predominately
associated with the ultraluminous IR QSOs found by \citet{36}.

Figure~\ref{classSEDs}
shows the {\it rest-frame} optical-to-far-IR broad-band SEDs
(at $b_J,\,r_F,\,J,\,H,\,K_s$, $12,\,25,\,60,\,100$ $\mu$m)
of this subsample of 2MASS AGN.
The SEDs are normalized to the rest-frame 12$\mu$m flux density since this
wavelength is expected to be a relatively unbiased indicator of AGN power
\citep[e.g.,][]{54}, i.e., independent of dust extinction
and star formation rate.
The rest-frame 12$\mu$m flux density
was estimated by linearly interpolating between the $12/(1+z)$
and $25/(1+z)$ wavelengths in the rest frame.
Figure~\ref{classSEDs} also shows the spread
(25$^{th}$ - 75$^{th}$ percentile range) observed for
optically selected PG QSOs. These lines are directly from \citet{15}.
The 2MASS red AGN all have higher normalized rest-frame far-IR emission
at $\lambda\gtrsim80\mu$m than the PG QSOs (at $>95\%$ significance).
Three 2MASS AGN also have higher
emission at rest-frame wavelengths $50\mu$m$<\lambda < 60\mu$m.
This indicates the far-IR emission in at least a handful of 2MASS AGN is
dominated by heated dust, either due to star formation, the central AGN or
both. This same dust could contribute to redenning of their optical-to-near-IR 
continua. It's important to note that the 2MASS AGN with far-IR
measurements are at the tail of distribution, i.e., just those with 
{\it IRAS} detections, and their IR SEDs are not necessarily
representative of 2MASS AGN in general. 

\subsection{Equivalent Width Distributions}

Figure~\ref{classBKvsEW} shows $b_J - K_s$ color as a function of  
equivalent width of the broad (permitted) emission line, H$\beta$,
and the narrow (forbidden) emission line, [O {\scriptsize III}]
($\lambda$ 5008\AA).
Traditionally, the broad emission
lines are thought to originate from
photoionized gas close to the central AGN known as
the broad line region (BLR), and the narrow lines from much
further out, to a few hundred parsecs or more, known as
the narrow line region (NLR).

Overall, the H$\beta$ and [O {\scriptsize III}] EW 
distributions for our type 1 AGN
are consistent with those of optically selected quasars (e.g., from SDSS).
However, there is a clear separation in these EWs
between our type 1 and 2 AGN and this is
consistent with the AGN unified model. The H$\beta$ line in most of the type 2 
AGN appears intrinsically weaker than that in type 1s. This is because 
fewer ionizing photons are reaching and being reprocessed by the
NLR, e.g., due to absorption and scattering by dust. 
In type 1s, we have a direct view of the BLR
line emission. Emission from the NLR can also contribute to
the observed H$\beta$ line flux in type 1s. 
The type 2 AGN also appear to be redder on average in $b_J - K_s$ than the 
type 1s. This is also consistent with the unified model if the 
continuum flux in type 2s suffers more extinction from dust, or if
obscuration of the central AGN allows us to see more contribution
from a ``redder'' host galaxy, e.g., from old stellar populations.

\placefigure{classBKvsEW}

The behavior in the EW of the narrow [O {\scriptsize III}] emission line
(Figure~\ref{classBKvsEW} - {\it right}) for type 1 and 2 AGN is reversed. 
Here, the [O {\scriptsize III}] emission from type 1s is reduced
relative to the optical continuum because the latter is
stronger (e.g., less extincted by dust) than that in type 2s.
This is consistent with the difference in $b_J - K_s$ 
color between the type 1 and 2 AGN being due to dust redenning.
Interestingly, $b_J - K_s$ color is weakly but significantly correlated 
with [O {\scriptsize III}] EW
for each of the type 1 and type 2 classes. Pearson's product-moment 
correlation coefficient is 0.29 and 0.32 
for the type 1s and type 2s respectively, with probabilities of
$<$0.1\% and $<$0.8\% of these measures occuring by chance.
Similar results are obtained using Kendall's non-parametric $\tau$ test.

These results suggest anisotropic obscuration of the central AGN 
and BLR, and the
bulk of the narrow-line emission originates from beyond  
the obscuring material. The type 2 AGN in particular require 
contamination by non-AGN light (e.g., galaxy hosts) if the
AGN continuum source is obscured. Otherwise their
[O {\scriptsize III}] EWs will be much larger than observed.
Scattered AGN light cannot account for the additional continuum since
it is expected to be less than a few percent in these
objects \citep[][]{35}.
Also, the overlap in EWs between
our type 1 AGN and optically selected QSOs  
implies host-galaxy contamination to their
optical continuum emission is minimal.

\section{Conclusions}
We have extended the 2MASS red AGN search to the southern equatorial sky and
increased the statistical base of red AGN by $\simeq35\%$ relative to previous 
surveys \citep[primarily from][]{7}.
We used the unique capabilities of the 6dF instrument and efficient
observing strategy of the 6dF Galaxy Redshift Survey to acquire spectra 
for 1182 candidates selected from the 2MASS PSC.
Our main conclusions are
\begin{enumerate}
\item{Classifications were secured for 432 spectra, of which 116 were
securely identified as type 1 AGN with broad permitted emission
lines exceeding 1000 $\kms$ (FWHM). 57 were identified as type 2 AGN,
all of which
are tentative due to the non-spectrophotometric nature of their spectra,
availability of appropriate emission line pairs,
and line flux measurement uncertainties.}

\item{A majority of the type 1 AGN are new, with only eight (or $\sim$6\%)
previously identified as AGN or QSOs in the literature. Most of them
were previously classified as galaxy-like and extended in the
optical (SuperCOSMOS digitized plates) or near-IR (2MASS Atlas Images).
95\% span the redshift range $0.05<z<0.5$.}

\item{Our selection method finds
a significantly higher fraction of local
galaxies containing AGN than in previous
blue-selected galaxy surveys. 
Our red ($J - K_s\,\gtrsim\,2$) color cut selects against 
intrinsically blue galaxy light.
We find that $\simeq24\%$ of our objects at $z < 0.05$ are
associated with an AGN.
For comparison, blue-selected galaxy samples are finding $\simeq4\%$.}

\item{Comparing our ratio of type 1 to type 2 AGN to that found in
previous studies of 2MASS red AGN, there is tentative
evidence that this ratio increases with a redder $J-K_s$ color cut.
This is consistent with the observation that most type 2 AGN have their
optical-to-near-IR emission dominated by blue galaxy light, and that
a redder $J-K_s$ cut will select more sources where the active nuclear
emission dominates, i.e., type 1 AGN.}

\item{The optical colors of the 2MASS AGN constrain any extinction by dust
to be $A_V<2$ mag relative to the SEDs of blue optically-selected QSOs.
Most of the type 1
AGN are too blue in their optical colors for dust to significantly
affect their optical/near-IR continua.
There are also a handful of red type 1 AGN 
(with $b_J-K_s\,\gtrsim\,4.5$) showing
excess far-IR emission at $\lambda\,\gtrsim\,80\mu$m, and some at
$50\mu$m$\,<\lambda < 60\mu$m. This suggests
at least some 2MASS AGN reside in dusty environments.}

\item{The distribution of H$\beta$ and [O {\scriptsize III}] equivalent widths,
and optical/near-IR colors
for type 1 and 2 AGN are consistent with AGN unified models.
The equivalent width of the [O {\scriptsize III}] emission line
correlates with $b_J - K_s$ color in both types of AGN,
suggesting anisotropic obscuration of the central AGN.
The type 2 AGN require a significant contribution to their optical
continua by stellar light to satisfy the equivement width measurements.}

\item{Overall, the optical properties of the 2MASS red AGN are not
dramatically different from those found in optical QSO samples.
This is most likely due to the relatively bright optical magnitude
limit imposed for reliable spectroscopic identification.
It appears that red AGN selection in 2MASS detects the most
luminous objects in the near-IR in the local universe to
$z\simeq0.6$. This is impetus to use similar methods for future deep
near-IR surveys. Indeed, such surveys have already begun to unravel
a new population of AGN at high redshift.}
\end{enumerate}

\acknowledgments

This publication makes use of data products from the Two Micron All Sky Survey,
which is a joint project of the University of Massachusetts and the Infrared
Processing and Analysis Center/California Institute of Technology, funded by 
the National Aeronautics and Space Administration and the National Science 
Foundation.
Funding for the SDSS and SDSS-II has been provided by the Alfred P. Sloan 
Foundation, the Participating Institutions, the National Science Foundation, 
the U.S. Department of Energy, the National Aeronautics and Space 
Administration, the Japanese Monbukagakusho, the Max Planck Society, and 
the Higher Education Funding Council for England. The SDSS Web Site 
is http://www.sdss.org/.
This research has made use of the NASA/IPAC Extragalactic Database (NED)
and the NASA/IPAC Infrared Science Archive (IRSA),
which are operated by the Jet Propulsion Laboratory, California Institute of 
Technology, under contract with the National Aeronautics and Space 
Administration.
The 6dFGS was carried out by the Anglo-Australian Observatory and the 6dFGS 
team\footnote{http://www.aao.gov.au/local/www/6df/6dFGSteam.html}, who have all 
contributed to the work presented here.
This work was carried out at the California Institute of Technology, with 
funding from the National Aeronautics and Space Administration.
RMC acknowledges support from the National Academy of Sciences
James Craig Watson Medal award. JPH was supported in part by NSF Grant 
AST0406906. We thank the anonymous referee for helpful comments.



{\it Facilities:} \facility{CTIO:2MASS}, \facility{UKST (6dF)}




\clearpage



\clearpage

\clearpage

\clearpage

\clearpage

\clearpage

\clearpage

\begin{figure}[h]
\begin{center}
{\bf Figure 7 available at:\\ http://web.ipac.caltech.edu/staff/fmasci/home/miscscience/f7.jpg}
\caption{Rest frame spectra of some new type 1 AGN overlayed
with emission lines typically found in QSO spectra. The
spectra for all objects listed in Tables~\ref{type1agn} and~\ref{type2agn}
can be viewed by querying the 6dF public database:
http://www.aao.gov.au/6dFGS/.}\label{spectra}
\end{center}
\end{figure}

\clearpage

\begin{figure}[h]
\begin{center}
\includegraphics[scale=0.5]{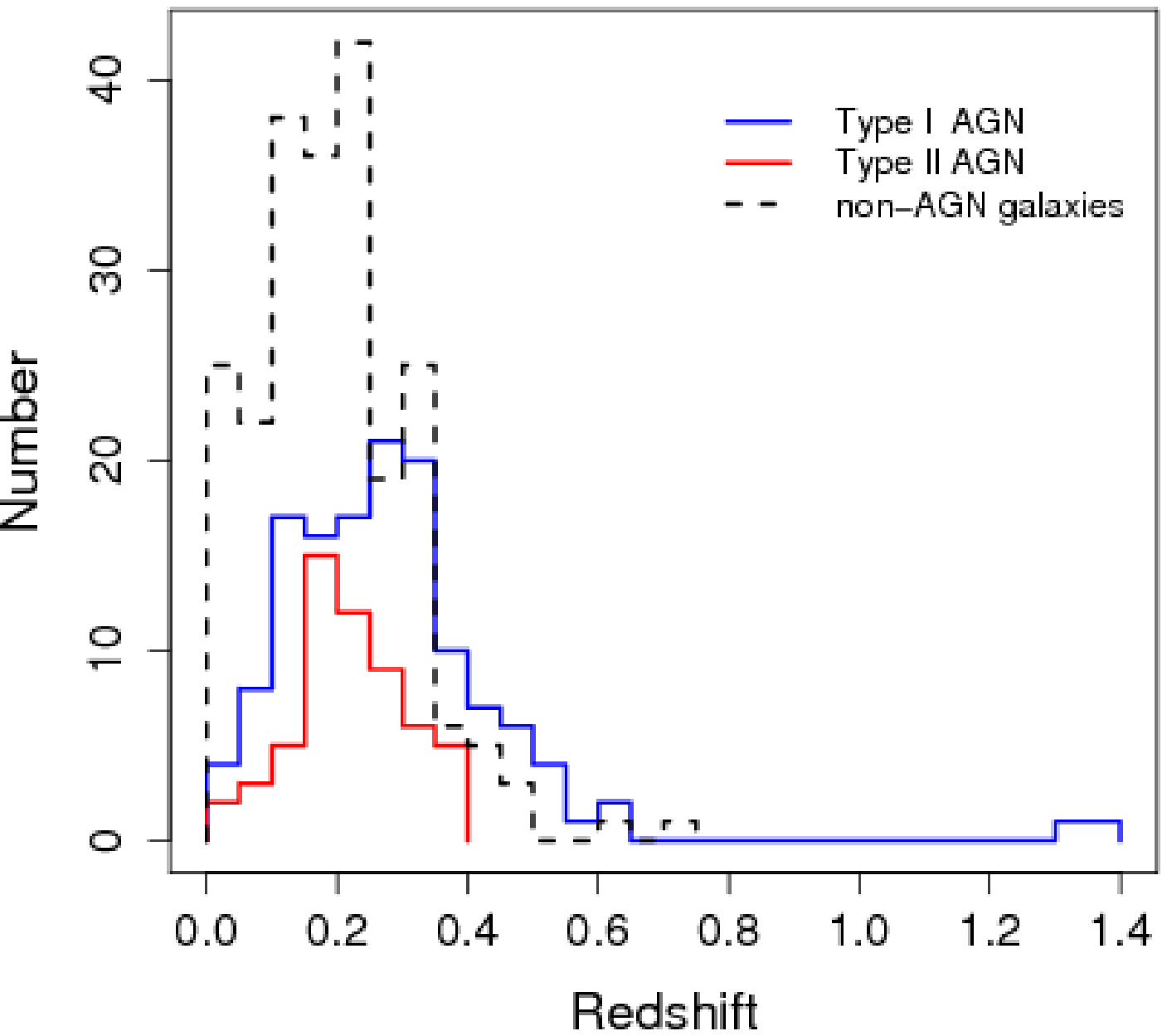}
\includegraphics[scale=0.5]{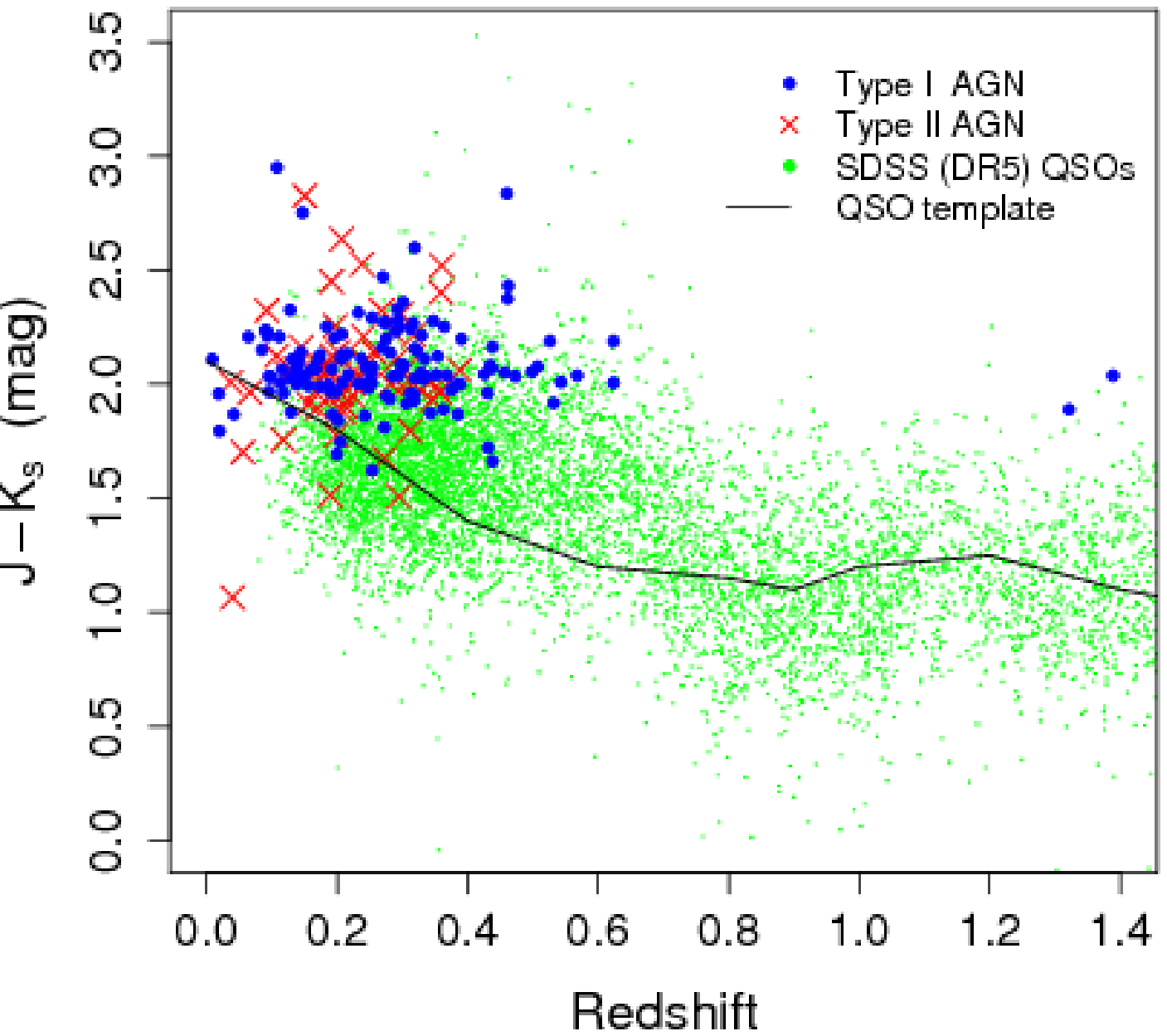}
\caption{({\it left}) Redshift distributions for our type 1 and 2 AGN,
and inactive galaxies with good quality spectra including
``unknown'' galaxy types.
({\it right}) $J - K_s$ color versus redshift
for our type 1 and 2 AGN, and SDSS QSOs.
The line shows the prediction for radio-quiet QSOs using the template
of \citet{10}. Uncertainties in $J-K_s$ are $\lesssim0.16$ mag
(1-$\sigma$).}\label{classHistoZ}
\end{center}
\end{figure}

\clearpage

\begin{figure}[h]
\begin{center}
\includegraphics[scale=0.5]{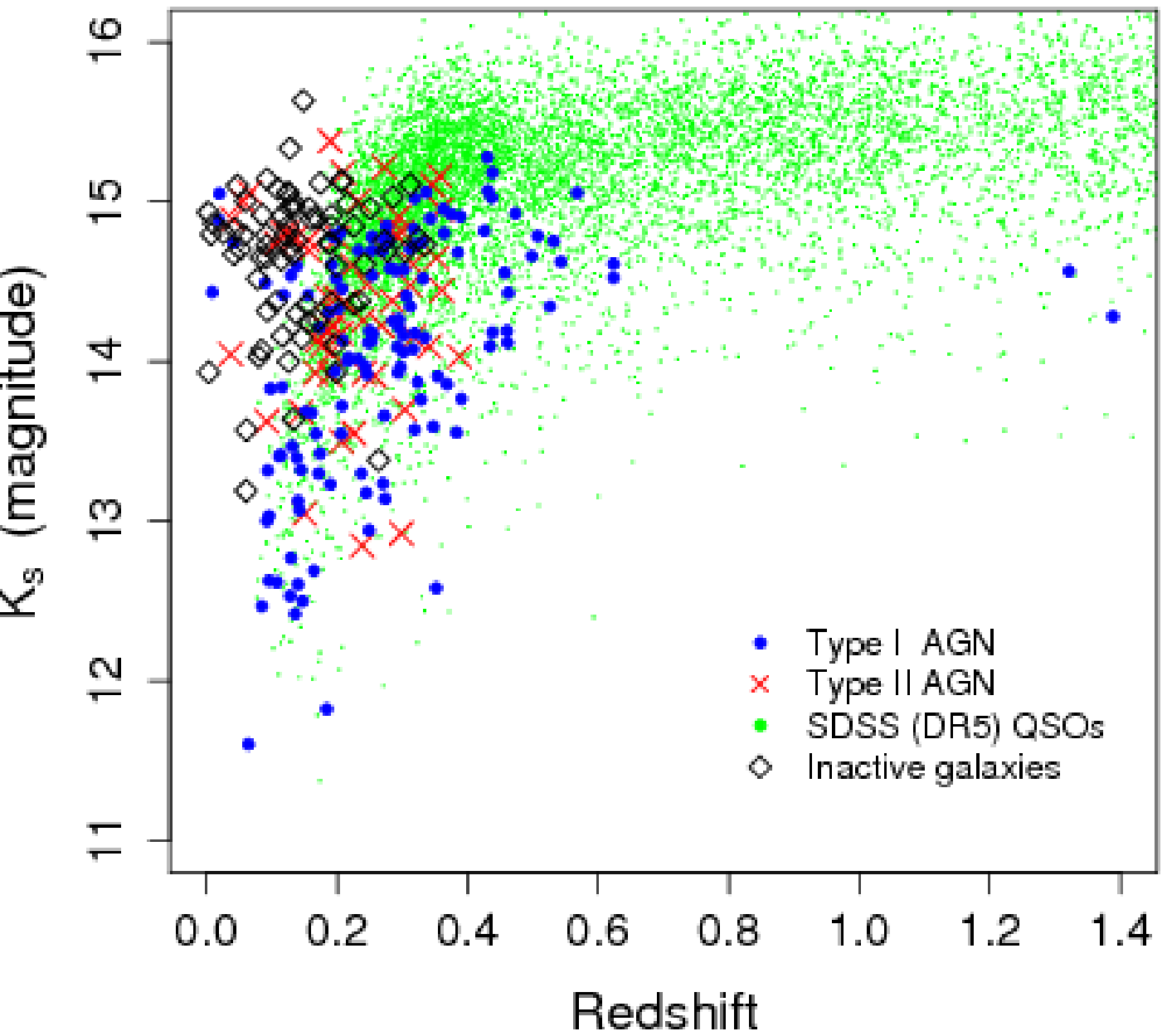}
\includegraphics[scale=0.5]{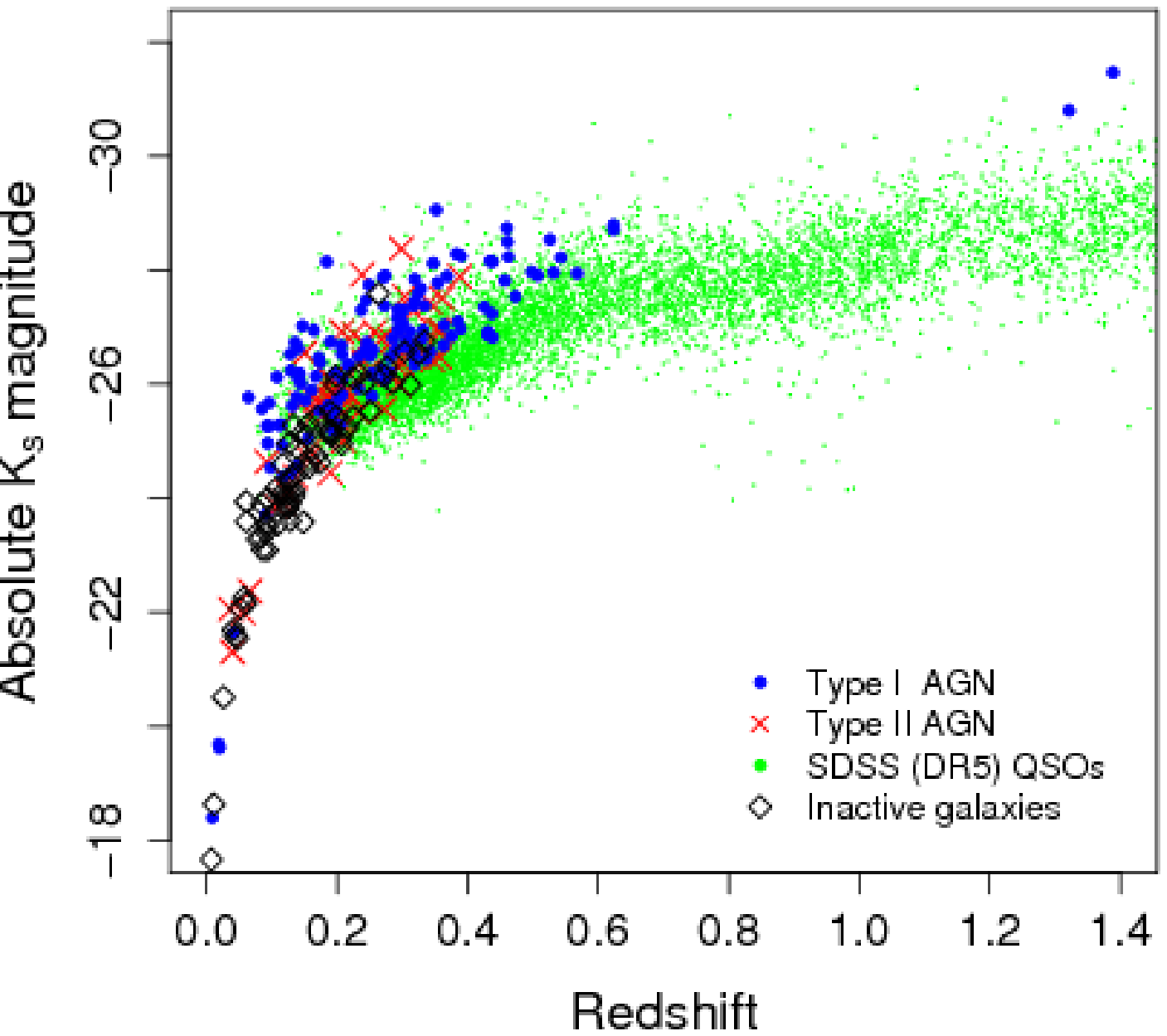}
\caption{Apparent $K_s$ magnitude ({\it left}) and absolute
$K_s$ magnitude ({\it right}) as a function
of redshift for 2MASS AGN, SDSS QSOs, and {\it securely}
identified inactive galaxies in our sample where
``unknown'' classifications were omitted.}\label{classKvsZ}
\end{center}
\end{figure}

\clearpage

\begin{figure}[h]
\begin{center}
\includegraphics[scale=0.5]{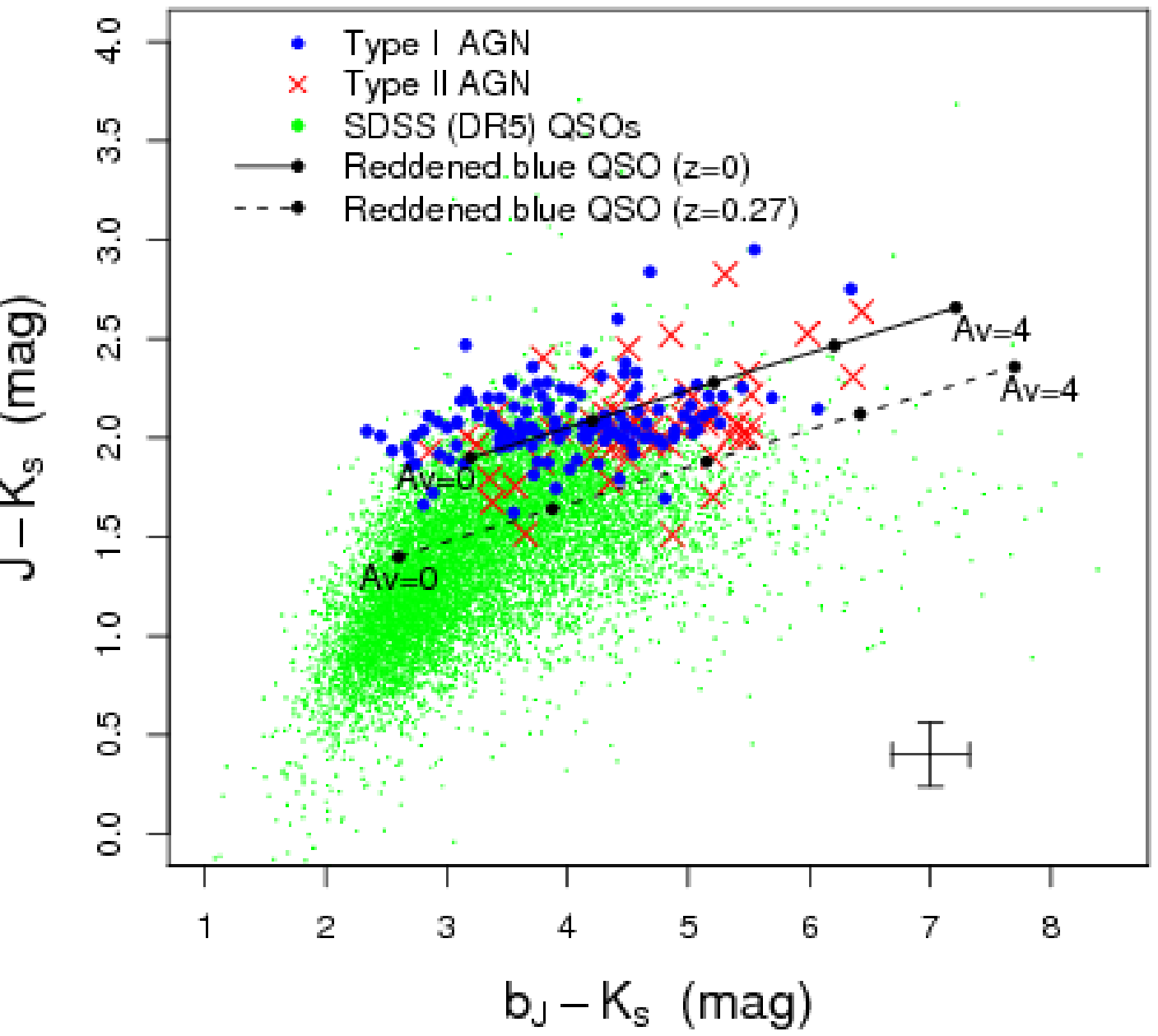}
\includegraphics[scale=0.5]{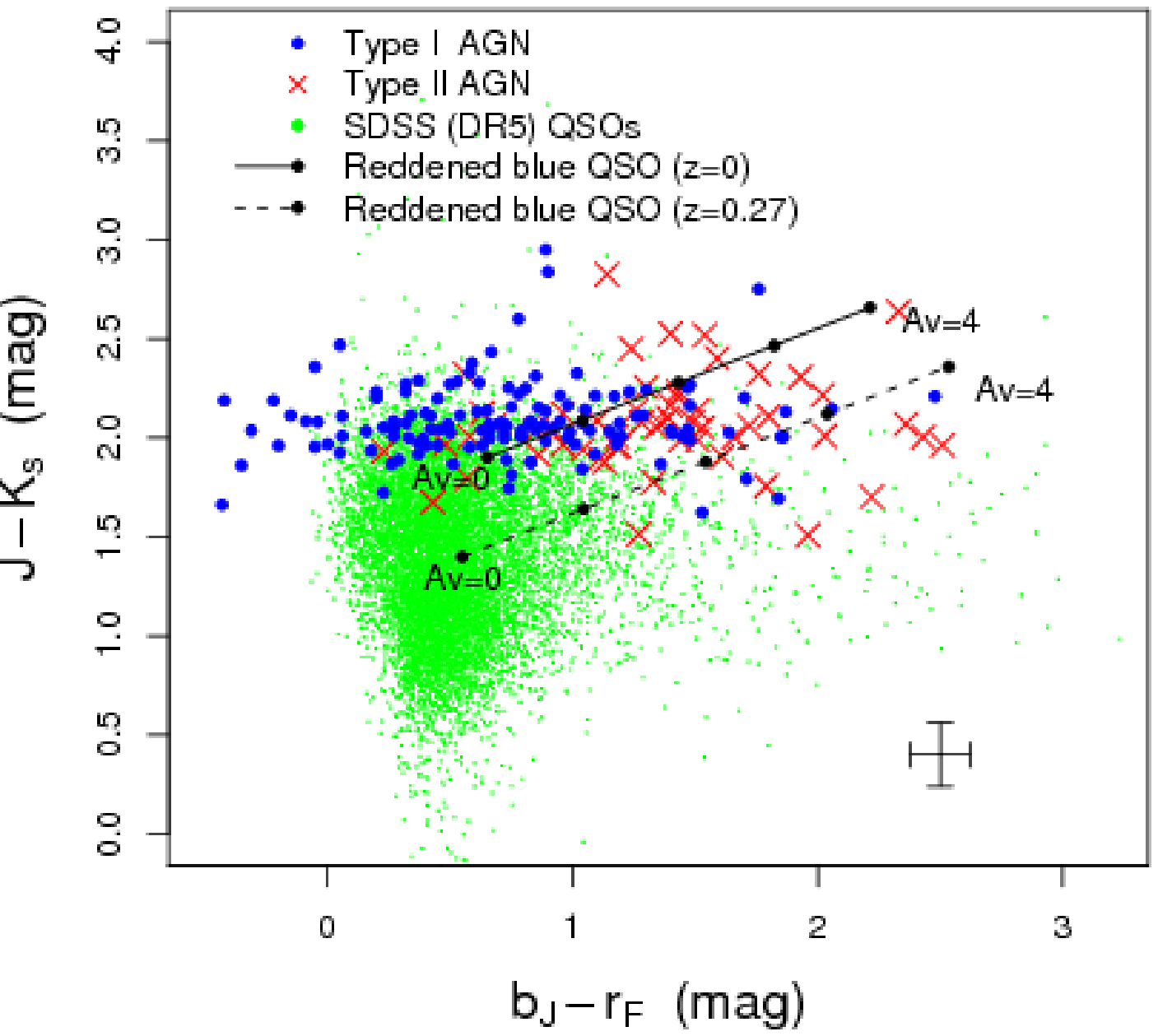}
\caption{Color-color plots involving $b_J,\,r_F,\,J,\,K_s$ for 2MASS red AGN
and SDSS QSOs. The lines with filled dots indicate changes in colors
due to pure dust reddening of a fiducial blue QSO
assuming a $1/\lambda$ extinction law at
$z=0$ (solid lines), and at the median redshift of our 
type 1 AGN, $z=0.27$ (dashed lines).
These predictions assume the intrinsic (unreddened) 
colors of a QSO from the
median composite of \citet{10}. The vertical/horizontal lines
denote (maximum) 1-$\sigma$ uncertainties along each axis:
$\simeq0.32$, $0.12$, and $0.16$ mag for $b_J-K_s$, $b_J-r_F$,
and $J-K_s$ respectively.}\label{classJKvsBK}
\end{center}
\end{figure}

\clearpage

\begin{figure}[h]
\begin{center}
\includegraphics[scale=1.0]{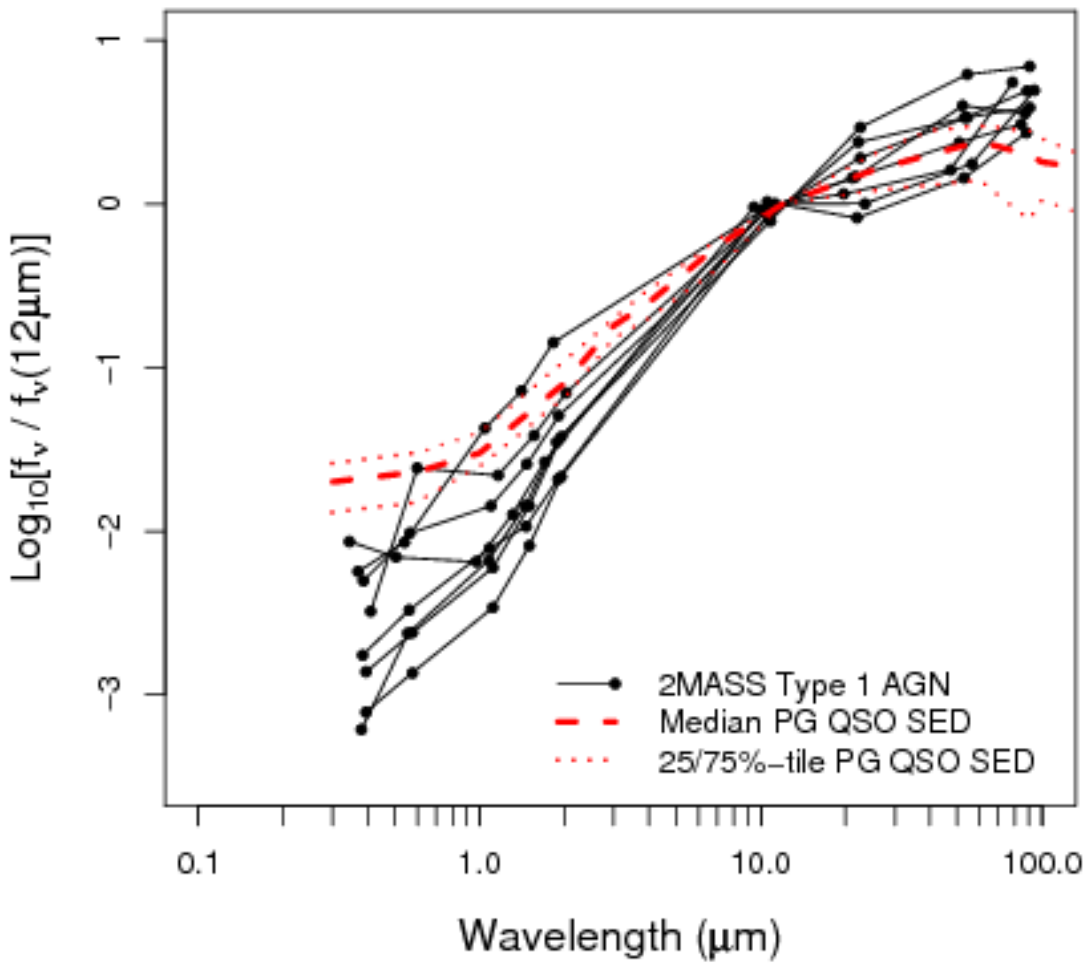}
\caption{Rest-frame SEDs of {\it IRAS} detected 2MASS type 1 AGN
(dots connected with lines) and the median, 25$^{th}$, and 75$^{th}$
percentiles of PG QSO SEDs from \citet[dashed and dotted lines]{15}.
These authors used survival statistics to account
for upper limits in the flux measurements of PG QSOs.
All SEDs are normalized to a linearly
interpolated rest-frame 12$\mu$m flux density (see text for details).
}\label{classSEDs}
\end{center}
\end{figure}

\clearpage

\begin{figure}[h]
\begin{center}
\includegraphics[scale=0.5]{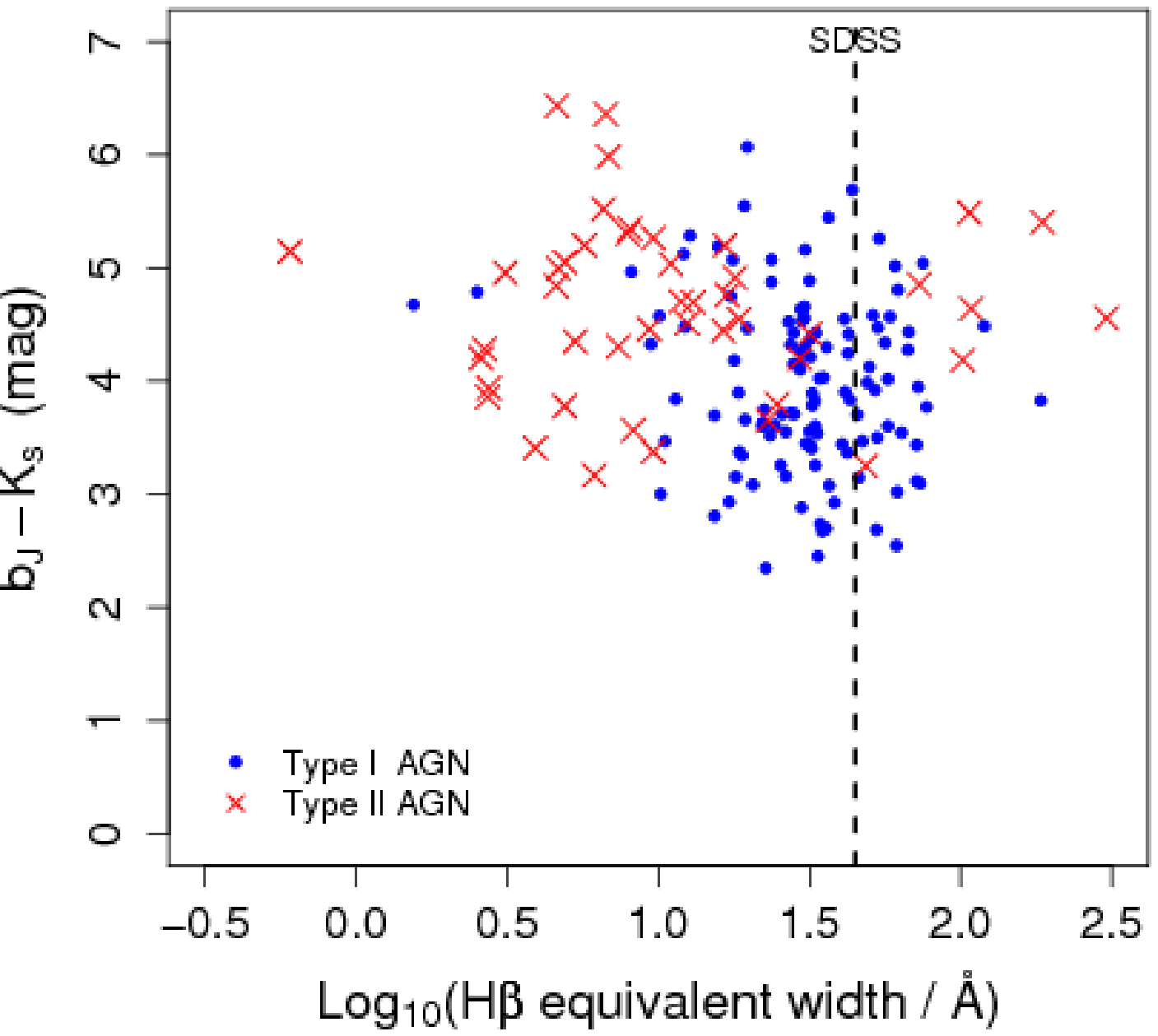}
\includegraphics[scale=0.5]{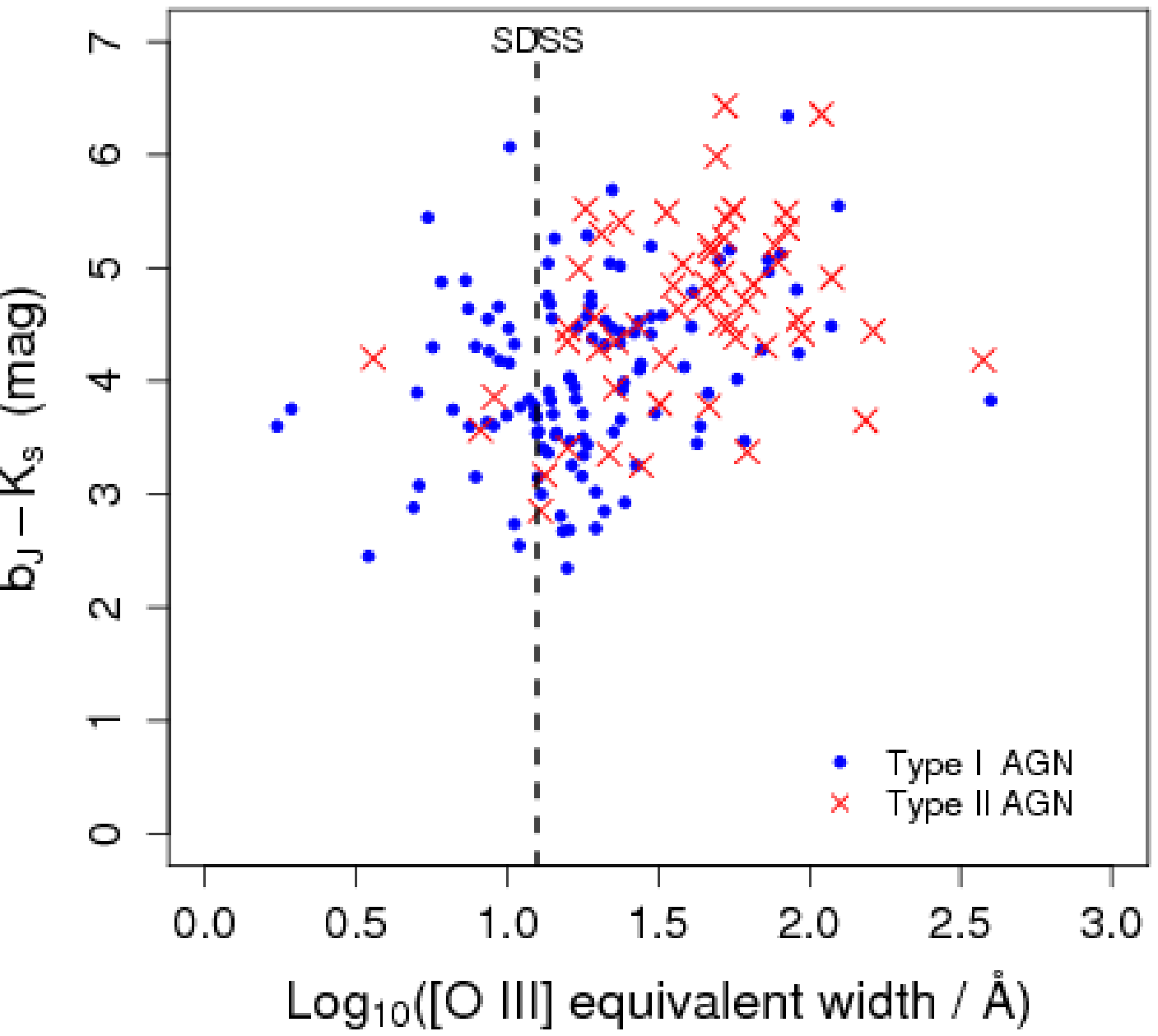}
\caption{$b_J - K_s$ color as a function of rest frame H$\beta$
equivalent width ({\it left}) and rest frame [O {\scriptsize III}]
equivalent width ({\it right}) for the 2MASS
type 1 and 2 AGN. Dashed vertical lines show equivalent
widths from the SDSS median composite spectrum of
\citet{56}. 1-$\sigma$ uncertainties are $\lesssim0.32$ mag in $b_J-K_s$
and $\lesssim0.18$ in $\log$$_{10}$(EW).}\label{classBKvsEW}
\end{center}
\end{figure}

\clearpage






\clearpage

\begin{table}
\begin{center}
\caption{Rest wavelength regions used in emission-line measurements.\label{lines}}
\begin{tabular}{lccc}
\tableline\tableline
Line & Center\tablenotemark{a} & Continuum Integration Limits & Line Integration Limits\\
 & (\AA) & (\AA) & (\AA)\\ 
\tableline
C {\scriptsize III}] & 1908.73 & 1800--1850, 1970--2010 & 1860--1955\\
Mg {\scriptsize II} & 2798.75 & 2650--2700, 3000--3050 & 2750--2850\\
{\rm [O {\scriptsize II}]} & 3728.48 & 3650--3700, 3770--3815 & 3710--3740\\
H$\beta$ & 4862.68 & 4740--4840, 4880--4940 & 4845--4880\\
{\rm [O {\scriptsize III}]} & 5008.24 & 4880--4940, 5030--5100 & 4990--5025\\
H$\alpha$ & 6564.61 & 6400--6520, 6630--6700 & 6550--6572\\
{\rm [N{\scriptsize II}]} & 6585.28 & 6400--6520, 6630--6700 & 6572--6594\\
{\rm [S{\scriptsize II}]} & 6725.48\tablenotemark{b} & 6630--6700, 6745--6845 & 6700--6745\\
\tableline
\end{tabular}
\tablenotetext{a}{As defined in \citet{56}}
\tablenotetext{b}{Average of doublet {\rm [S{\scriptsize II}]} $\lambda\lambda 6718.29,6732.67$}
\end{center}
\end{table}

\clearpage

\begin{deluxetable}{lcccc}
\tabletypesize{\small}
\rotate
\tablecaption{Source classifications and comparisons to previous studies.\label{summ}}
\tablewidth{0pt}
\tablehead{
\colhead{} &
\colhead{This Study} &
\colhead{\citet{7}} &
\colhead{\citet{12}} &
\colhead{\citet{12}} \\
\colhead{} &
\colhead{($J-K_s\gtrsim1.7$)} &
\colhead{($J-K_s>2.0$)} &
\colhead{($J-K_s>1.2$)} & 
\colhead{($J-K_s>1.8$)}
}
\startdata
Number Classified & 432 & 664 & 1304 & 66\\
Stars (\%) & 23 (5.3)\tablenotemark{a} & 66 (9.9) & 330 (25.3) & 6 (9.1)\\
Unknown Galaxies (\%) & 193 (44.7) & \ldots & \ldots & \ldots\\
Early-type Galaxies (\%) & 17 (3.9) & \ldots & \ldots & \ldots\\
SB/Late-type Galaxies (\%) & 26\tablenotemark{c} (6.0) & \ldots & 106 (8.1) & \ldots \\
Type 1 AGN (\%) & 116\tablenotemark{b} (26.8) & 385 (57.9) & 14 (1.1) & 4 (6.0)\\
Type 2 AGN (\%) & 57\tablenotemark{c} (13.2) & 100 (15.0) & 23 (1.8) & 0 (0.0)\\
\enddata
\tablenotetext{a}{Values in parenthesis are percentages of the
                  ``Number Classified'' (first row)}
\tablenotetext{b}{The 20 probable identifications are excluded here and classified as
                  ``Unknown Galaxies''}
\tablenotetext{c}{We declare all these to be probable identifications}
\end{deluxetable}

\clearpage

\begin{deluxetable}{lccccccccl}
\tabletypesize{\scriptsize}
\rotate
\tablecaption{Type 1 AGN\label{type1agn}}
\tablewidth{0pt}
\tablehead{
\colhead{Name} &
\colhead{R.A.\tablenotemark{a}} & 
\colhead{Dec.\tablenotemark{a}} & 
\colhead{$b_J$} & 
\colhead{$r_F$} & 
\colhead{$K$} & 
\colhead{$J-K_s$} & 
\colhead{Redshift} & 
\colhead{ID\tablenotemark{b}} &
\colhead{Other Name}
}
\startdata
2MASS J00004025-0541009 & 00 00 40.26 & -05 41 00.9 & 17.84 & 16.750 & 13.315 & 2.212 & 0.094 & T1 & 2MASX J00004028-0541012\\
2MASS J00020731-4722504 & 00 02 07.32 & -47 22 50.5 & 18.52 & 18.190 & 14.909 & 1.999 & 0.389 & T1 &\\
2MASS J00085227-6233137 & 00 08 52.30 & -62 33 15.3 & 17.27 & 16.530 & 11.825 & 2.252 & 0.184 & T1 & IRAS F00063-6249\\
2MASS J00114353-5033299 & 00 11 43.54 & -50 33 30.0 & 18.09 & 16.800 & 13.125 & 2.111 & 0.140 & T1 & 2MASX J00114350-5033302\\
2MASS J00163310-0542315 & 00 16 33.10 & -05 42 31.4 & 17.61 & 17.070 & 13.299 & 2.113 & 0.237 & T1 & 1RXS J001633.2-054227\\
2MASS J00202646-0254583 & 00 20 26.46 & -02 54 58.2 & 17.97 & 17.680 & 14.951 & 1.889 & 0.363 & T1 &\\
2MASS J00265121-0159238 & 00 26 51.16 & -01 59 23.7 & 19.36 & 17.720 & 15.020 & 2.026 & 0.319 & T1 &\\
2MASS J00315009-3000482 & 00 31 50.08 & -30 00 48.3 & 17.46 & 17.190 & 14.094 & 2.082 & 0.435 & T1 &\\
2MASS J00345758-3221311 & 00 34 57.58 & -32 21 31.2 & 18.77 & 17.590 & 14.303 & 1.997 & 0.186 & T1 & 2dFGRS S440Z049\\
2MASS J00470010-1747025 & 00 47 00.05 & -17 47 02.0 & 18.84 & 17.080 & 12.500 & 2.752 & 0.147 & T1 & IRAS 00444-1803\\
2MASS J00573811-1406173 & 00 57 38.12 & -14 06 17.3 & 18.56 & 17.780 & 14.458 & 2.219 & 0.207 & T1 &\\
2MASS J00575201-3329048 & 00 57 52.03 & -33 29 04.6 & 18.53 & 17.550 & 14.691 & 2.162 & 0.270 & T1 &\\
2MASS J00593208-1540302 & 00 59 32.11 & -15 40 29.7 & 18.33 & 16.800 & 14.778 & 1.620 & 0.254 & T1 & 1RXS J005932.7-154032\\
2MASS J01144251-5136137 & 01 14 42.46 & -51 36 13.8 & 18.71 & 17.970 & 14.806 & 1.746 & 0.205 & T1 &\\
2MASS J01274593-4453170 & 01 27 45.95 & -44 53 17.0 & 17.83 & 17.320 & 14.683 & 1.868 & 0.385 & T1 &\\
2MASS J01283395-2358359 & 01 28 33.96 & -23 58 35.8 & 18.58 & 17.910 & 14.428 & 2.433 & 0.462 & PT1 &\\
2MASS J01290719-2356300 & 01 29 07.24 & -23 56 30.2 & 19.48 & 17.770 & 15.053 & 1.793 & 0.020 & T1 & APMUKS B012644.60-241200.3\\
2MASS J01340171-5618454 & 01 34 01.70 & -56 18 45.5 & 17.46 & 17.190 & 13.762 & 2.050 & 0.328 & PT1 & SUMSS J013401-561844\\
2MASS J01354636-3539148 & 01 35 46.35 & -35 39 15.3 & 18.03 & 17.050 & 13.396 & 2.000 & 0.137 & T1 & 2MASX J01354638-3539151\\
2MASS J01380953-0109201 & 01 38 09.54 & -01 09 20.1 & 16.68 & 16.360 & 13.141 & 2.271 & 0.273 & T1 & SDSS J013809.53-010920.2\\
2MASS J01432356-0124523 & 01 43 23.57 & -01 24 52.4 & 18.04 & 17.250 & 13.858 & 2.036 & 0.368 & T1 &\\
2MASS J01553002-0857040 & 01 55 30.03 & -08 57 04.0 & 17.95 & 16.910 & 12.690 & 2.072 & 0.164 & T1 & SDSS J015530.02-085704.0\\
2MASS J01593068-1128584 & 01 59 30.69 & -11 28 58.5 & 18.15 & 17.270 & 13.676 & 2.062 & 0.161 & T1 &\\
2MASS J01593518-2010583 & 01 59 35.20 & -20 10 58.4 & 19.15 & 17.670 & 14.078 & 2.267 & 0.317 & PT1 & APMUKS B015713.48-202531.4\\
2MASS J02112824-0640013 & 02 11 28.22 & -06 40 00.8 & 18.42 & 18.000 & 14.816 & 2.045 & 0.426 & PT1 &\\
2MASS J02120146-0201539 & 02 12 01.47 & -02 01 53.8 & 19.20 & 17.720 & 14.182 & 2.163 & 0.438 & T1 &\\
2MASS J02144017-6839338 & 02 14 40.21 & -68 39 33.8 & 18.96 & 17.870 & 14.412 & 1.916 & 0.305 & PT1 &\\
2MASS J02155303-4709570 & 02 15 53.04 & -47 09 57.1 & 17.64 & 16.690 & 12.602 & 2.085 & 0.140 & T1 & 2MASX J02155306-4709573\\
2MASS J02181461-1012491 & 02 18 14.61 & -10 12 49.1 & 17.86 & 18.170 & 15.054 & 2.036 & 0.568 & PT1 &\\
2MASS J02221945-5332391 & 02 22 19.44 & -53 32 39.1 & 16.61 & 15.540 & 12.626 & 2.035 & 0.095 & T1 &\\
2MASS J02252283-2441528 & 02 25 22.84 & -24 41 52.8 & 17.81 & 17.390 & 14.435 & 2.109 & 0.009 & T1 &\\
2MASS J02254799-1429454 & 02 25 48.01 & -14 29 45.4 & 18.63 & 17.930 & 15.026 & 2.068 & 0.437 & T1 &\\
2MASS J02255003-0601450 & 02 25 50.04 & -06 01 45.1 & 17.13 & 16.380 & 13.574 & 2.156 & 0.319 & T1 &\\
2MASS J02270655-3220448 & 02 27 06.52 & -32 20 44.7 & 18.42 & 17.530 & 14.119 & 1.985 & 0.249 & T1 & 2dFGRS S463Z081\\
2MASS J02293513-0552094 & 02 29 35.14 & -05 52 09.4 & 17.47 & 16.940 & 13.928 & 2.280 & 0.293 & T1 &\\
2MASS J02313792-2308312 & 02 31 37.93 & -23 08 31.2 & 15.68 & 15.410 & 12.941 & 2.008 & 0.249 & T1 &\\
2MASS J02331354-1506540 & 02 33 13.55 & -15 06 54.1 & 18.69 & 18.060 & 15.059 & 2.035 & 0.337 & PT1 &\\
2MASS J02365236-5554463 & 02 36 52.33 & -55 54 46.3 & 17.42 & 17.620 & 14.342 & 1.960 & 0.312 & T1 &\\
2MASS J02391868-0115211 & 02 39 18.69 & -01 15 21.1 & 18.52 & 18.150 & 14.919 & 1.975 & 0.374 & T1 & SDSS J023918.70-011521.0\\
2MASS J02452486-1623432 & 02 45 24.86 & -16 23 43.3 & 18.81 & 17.660 & 14.252 & 2.034 & 0.286 & T1 & APMUKS B024304.02-163618.0\\
2MASS J02460800-1132367 & 02 46 08.01 & -11 32 36.6 & 16.39 & 16.340 & 13.234 & 2.470 & 0.270 & T1 & IRAS F02437-1145\\
2MASS J02535862-4558236 & 02 53 58.63 & -45 58 23.9 & 18.87 & 17.970 & 14.189 & 2.837 & 0.460 & PT1 &\\
2MASS J02564173-1231478 & 02 56 41.72 & -12 31 48.4 & 17.27 & 17.110 & 14.926 & 2.034 & 0.473 & T1 &\\
2MASS J02574845-0918443 & 02 57 48.45 & -09 18 44.6 & 18.07 & 17.020 & 13.319 & 2.140 & 0.145 & T1 & 2MASX J02574849-0918440\\
2MASS J02593835-1513092 & 02 59 38.35 & -15 13 09.2 & 17.16 & 16.760 & 13.906 & 2.122 & 0.354 & T1 & NVSS J025938-151306\\
2MASS J03012933-1632400 & 03 01 29.35 & -16 32 39.8 & 17.00 & 16.940 & 14.150 & 2.110 & 0.334 & T1 &\\
2MASS J03103479-6412052 & 03 10 34.80 & -64 12 05.2 & 17.76 & 16.750 & 13.232 & 1.967 & 0.190 & T1 & 2MASX J03103473-6412054\\
2MASS J03153812-0912337 & 03 15 38.11 & -09 12 34.0 & 16.29 & 15.440 & 12.580 & 2.041 & 0.352 & T1 &\\
2MASS J03173580-3817251 & 03 17 35.81 & -38 17 25.3 & 18.21 & 17.840 & 14.693 & 2.290 & 0.254 & T1 &\\
2MASS J03232133-2557442 & 03 23 21.34 & -25 57 44.3 & 17.73 & 17.530 & 14.568 & 2.232 & 0.290 & T1 &\\
2MASS J03261346-2018137 & 03 26 13.45 & -20 18 13.7 & 18.41 & 17.760 & 14.582 & 2.139 & 0.281 & PT1 &\\
2MASS J03360954-0619128 & 03 36 09.52 & -06 19 12.9 & 17.99 & 18.420 & 15.181 & 1.662 & 0.438 & T1 & SDSS J033609.54-061913.1\\
2MASS J03374281-2522095 & 03 37 42.81 & -25 22 09.5 & 17.41 & 16.910 & 13.660 & 2.270 & 0.273 & T1 &\\
2MASS J03412733-0906436 & 03 41 27.33 & -09 06 43.5 & 16.99 & 16.330 & 13.554 & 1.997 & 0.383 & T1 &\\
2MASS J03413408-0547154 & 03 41 34.11 & -05 47 15.4 & 18.37 & 17.170 & 14.595 & 2.007 & 0.137 & T1 & SDSS J034134.07-054715.7\\
2MASS J03440463-1252226 & 03 44 04.62 & -12 52 22.6 & 17.93 & 16.700 & 13.001 & 2.230 & 0.093 & T1 & 2MASX J03440465-1252222\\
2MASS J03561630-1237095 & 03 56 16.33 & -12 37 09.2 & 17.50 & 17.550 & 14.828 & 1.955 & 0.318 & T1 & 1RXS J035616.4-123712\\
2MASS J03561996-6251391 & 03 56 19.96 & -62 51 39.1 & 18.16 & 17.270 & 12.617 & 2.948 & 0.108 & T1 & 2MASX J03561995-6251391\\
2MASS J04023421-3152594 & 04 02 34.22 & -31 52 59.2 & 17.86 & 17.030 & 13.963 & 2.085 & 0.297 & PT1 & NVSS J040234-315301\\
2MASS J04043381-4039545 & 04 04 33.81 & -40 39 54.6 & 18.65 & 17.860 & 14.525 & 2.006 & 0.623 & T1 &\\
2MASS J04170175-0309594 & 04 17 01.74 & -03 09 59.7 & 18.53 & 17.580 & 14.745 & 2.214 & 0.330 & T1 &\\
2MASS J04225656-1854422 & 04 22 56.56 & -18 54 42.3 & 16.89 & 14.410 & 11.605 & 2.209 & 0.064 & T1 & 2MASX J04225654-1854424\\
2MASS J04243795-2318335 & 04 24 37.96 & -23 18 33.6 & 18.14 & 17.350 & 14.621 & 2.009 & 0.542 & T1 &\\
2MASS J04324870-0921117 & 04 32 48.70 & -09 21 11.6 & 18.58 & 17.690 & 14.000 & 2.133 & 0.216 & T1 &\\
2MASS J04332250-1422287 & 04 33 22.51 & -14 22 28.6 & 17.42 & 16.800 & 13.592 & 2.276 & 0.347 & T1 & 1WGA J0433.3-1422\\
2MASS J04352649-1643460 & 04 35 26.50 & -16 43 46.0 & 17.30 & 16.310 & 13.828 & 2.031 & 0.098 & T1 &\\
2MASS J04360031-0405540 & 04 36 00.32 & -04 05 54.0 & 18.15 & 17.830 & 14.492 & 2.235 & 0.091 & T1 &\\
2MASS J04364839-1123559 & 04 36 48.40 & -11 23 55.9 & 17.38 & 16.770 & 13.720 & 2.129 & 0.207 & T1 & NVSS J043647-112347\\
2MASS J04411070-0639383 & 04 41 10.71 & -06 39 38.3 & 18.73 & 17.430 & 14.706 & 2.241 & 0.312 & T1 &\\
2MASS J04433082-3508499 & 04 43 30.78 & -35 08 49.9 & 18.24 & 17.970 & 14.553 & 2.049 & 0.457 & T1 &\\
2MASS J04481925-2158473 & 04 48 19.25 & -21 58 47.1 & 18.57 & 17.400 & 13.408 & 2.209 & 0.111 & T1 & 2MASX J04481922-2158468\\
2MASS J05022413-3546422 & 05 02 24.15 & -35 46 42.3 & 18.34 & 17.950 & 14.417 & 1.993 & 0.155 & T1 &\\
2MASS J05043443-1521193 & 05 04 34.42 & -15 21 19.3 & 17.21 & 16.760 & 13.768 & 2.200 & 0.390 & T1 &\\
2MASS J05063975-3416422 & 05 06 39.77 & -34 16 42.2 & 18.96 & 17.480 & 14.211 & 1.990 & 0.173 & T1 & APMUKS B050450.33-342037.7\\
2MASS J05222203-4524370 & 05 22 21.99 & -45 24 36.8 & 17.73 & 17.390 & 13.296 & 2.112 & 0.172 & T1 & 1WGA J0522.3-4524\\
2MASS J05372445-3817586 & 05 37 24.48 & -38 17 58.7 & 18.02 & 17.300 & 14.517 & 2.023 & 0.332 & T1 &\\
2MASS J10012986-0338334 & 10 01 29.88 & -03 38 33.5 & 17.75 & 17.090 & 14.280 & 2.038 & 1.389 & PT1 &\\
2MASS J10264404-0425408 & 10 26 44.05 & -04 25 40.6 & 16.99 & 16.930 & 14.539 & 2.007 & 0.254 & T1 &\\
2MASS J10443748-0705162 & 10 44 37.49 & -07 05 15.9 & 17.25 & 17.340 & 14.163 & 2.085 & 0.301 & PT1 &\\
2MASS J11291836-2855531 & 11 29 18.36 & -28 55 53.1 & 18.87 & 18.050 & 14.604 & 2.066 & 0.192 & T1 &\\
2MASS J11294994-1153488 & 11 29 49.93 & -11 53 49.4 & 17.95 & 17.720 & 15.068 & 1.721 & 0.431 & T1 &\\
2MASS J12101994-1157105 & 12 10 19.91 & -11 57 10.4 & 16.80 & 16.950 & 13.544 & 2.111 & 0.206 & T1 & LCRS B120745.5-114029\\
2MASS J12140669-2948164 & 12 14 06.69 & -29 48 16.4 & 17.66 & 17.430 & 14.658 & 2.053 & 0.498 & T1 &\\
2MASS J12154013-2350122 & 12 15 40.14 & -23 50 12.2 & 18.89 & 17.420 & 14.015 & 2.001 & 0.232 & T1 & 2MASX J12154013-2350126\\
2MASS J12524011-3036579 & 12 52 40.12 & -30 36 58.0 & 18.75 & 18.350 & 15.278 & 1.962 & 0.430 & PT1 &\\
2MASS J12590021-3152245 & 12 59 00.21 & -31 52 24.5 & 18.66 & 18.080 & 14.090 & 2.327 & 0.292 & T1 &\\
2MASS J13210070-0806482 & 13 21 00.72 & -08 06 47.9 & 17.88 & 17.560 & 14.786 & 2.074 & 0.507 & T1 &\\
2MASS J13240925-1525097 & 13 24 09.26 & -15 25 09.4 & 17.69 & 17.320 & 14.754 & 1.917 & 0.531 & PT1 &\\
2MASS J13265209-1506385 & 13 26 52.11 & -15 06 38.8 & 19.94 & 17.880 & 13.872 & 2.144 & 0.323 & T1 &\\
2MASS J13450643-2705409 & 13 45 06.44 & -27 05 41.0 & 18.22 & 17.640 & 14.783 & 1.950 & 0.276 & T1 &\\
2MASS J13500251-2619308 & 13 50 02.52 & -26 19 30.8 & 18.46 & 17.010 & 14.133 & 2.006 & 0.208 & T1 & 2MASX J13500249-2619306\\
2MASS J14403905-1927129 & 14 40 39.05 & -19 27 13.0 & 18.21 & 17.400 & 14.264 & 2.251 & 0.295 & T1 &\\
2MASS J15400255-1451363 & 15 40 02.54 & -14 51 36.3 & 19.03 & 17.670 & 14.783 & 1.868 & 0.193 & T1 &\\
2MASS J15470139-0632099 & 15 47 01.41 & -06 32 09.9 & 18.34 & 17.420 & 14.127 & 2.048 & 0.252 & T1 &\\
2MASS J15475594-0610480 & 15 47 55.95 & -06 10 48.3 & 18.29 & 17.460 & 14.539 & 1.877 & 0.129 & T1 &\\
2MASS J20202833-3014140 & 20 20 28.33 & -30 14 14.2 & 19.63 & 17.930 & 13.939 & 2.204 & 0.196 & T1 &\\
2MASS J20223079-3236086 & 20 22 30.79 & -32 36 08.6 & 17.45 & 16.950 & 13.069 & 2.030 & 0.142 & T1 &\\
2MASS J20331662-2253169 & 20 33 16.62 & -22 53 17.0 & 18.59 & 17.330 & 13.467 & 2.109 & 0.131 & T1 & PKS 2030-23\\
2MASS J20344129-5126573 & 20 34 41.28 & -51 26 57.3 & 17.81 & 16.400 & 12.772 & 2.024 & 0.128 & T1 & 2MASX J20344130-5126572\\
2MASS J20345400-3131563 & 20 34 53.99 & -31 31 56.4 & 18.35 & 17.310 & 14.332 & 1.842 & 0.202 & T1 &\\
2MASS J20400728-5237227 & 20 40 07.30 & -52 37 22.6 & 17.00 & 15.980 & 12.532 & 2.325 & 0.128 & T1 & 2MASX J20400731-5237224\\
2MASS J20465715-7400040 & 20 46 57.15 & -74 00 04.1 & 17.46 & 17.680 & 14.345 & 2.189 & 0.526 & T1 & PKS 2041-741\\
2MASS J20585386-2409418 & 20 58 53.86 & -24 09 41.8 & 19.10 & 17.700 & 14.212 & 2.038 & 0.294 & PT1 &\\
2MASS J21071766-5503097 & 21 07 17.66 & -55 03 10.1 & 18.60 & 18.010 & 14.117 & 2.374 & 0.461 & PT1 & SUMSS J210717-550308\\
2MASS J21132600-6316184 & 21 13 26.04 & -63 16 18.4 & 18.49 & 17.780 & 14.172 & 2.075 & 0.255 & T1 &\\
2MASS J21133234-6257134 & 21 13 32.35 & -62 57 13.5 & 17.67 & 17.010 & 13.545 & 2.066 & 0.168 & T1 &\\
2MASS J21260042-2558396 & 21 26 00.42 & -25 58 39.6 & 18.73 & 17.900 & 14.894 & 1.873 & 0.343 & T1 &\\
2MASS J21333132-6120301 & 21 33 31.31 & -61 20 30.2 & 18.32 & 17.590 & 13.840 & 2.062 & 0.116 & T1 &\\
2MASS J21570360-7322204 & 21 57 03.61 & -73 22 20.5 & 17.82 & 16.640 & 13.033 & 1.967 & 0.096 & T1 & 2MASX J21570361-7322205\\
2MASS J21571362-4201497 & 21 57 13.61 & -42 01 49.2 & 18.64 & 17.910 & 14.562 & 1.888 & 1.321 & T1 & LCRS B215408.7-421607\\
2MASS J21572015-4036138 & 21 57 20.17 & -40 36 13.8 & 17.77 & 17.820 & 14.058 & 2.359 & 0.301 & T1 &\\
2MASS J21572316-3123064 & 21 57 23.17 & -31 23 06.5 & 16.50 & 15.640 & 12.469 & 2.151 & 0.085 & T1 & 1RXS J215723.3-31230\\
2MASS J22161121-3947337 & 22 16 11.23 & -39 47 33.7 & 18.35 & 16.500 & 13.920 & 2.004 & 0.246 & T1 &\\
2MASS J22165321-4451568 & 22 16 53.21 & -44 51 57.0 & 15.83 & 15.340 & 12.421 & 2.067 & 0.135 & T1 & LCRS B221349.8-450657\\
2MASS J22170559-3619460 & 22 17 05.60 & -36 19 46.1 & 17.58 & 17.120 & 13.687 & 2.040 & 0.151 & T1 &\\
2MASS J22180018-3957228 & 22 18 00.19 & -39 57 22.7 & 17.83 & 16.640 & 13.177 & 2.072 & 0.244 & T1 &\\
2MASS J22212645-4757598 & 22 21 26.46 & -47 57 59.6 & 17.97 & 17.970 & 14.520 & 1.969 & 0.198 & T1 &\\
2MASS J22250207-1841401 & 22 25 02.09 & -18 41 40.0 & 17.50 & 17.540 & 14.573 & 2.080 & 0.302 & T1 &\\
2MASS J22253449-1939566 & 22 25 34.50 & -19 39 56.6 & 19.38 & 17.910 & 14.805 & 2.253 & 0.364 & PT1 &\\
2MASS J22385748-0539206 & 22 38 57.49 & -05 39 20.6 & 18.62 & 16.750 & 13.425 & 2.130 & 0.174 & T1 & 1WGA J2238.9-0539\\
2MASS J22423001-2745299 & 22 42 30.02 & -27 45 29.6 & 19.58 & 17.740 & 14.773 & 1.695 & 0.199 & T1 &\\
2MASS J22431660-2020570 & 22 43 16.59 & -20 20 57.1 & 18.19 & 17.990 & 14.849 & 2.202 & 0.276 & T1 &\\
2MASS J22433493-3928397 & 22 43 34.94 & -39 28 39.6 & 19.10 & 17.630 & 14.025 & 2.042 & 0.217 & T1 & APMUKS B224042.73-394424.8\\
2MASS J22553836-0245335 & 22 55 38.34 & -02 45 33.1 & 18.48 & 17.730 & 14.764 & 1.812 & 0.273 & T1 & 1WGA J2255.6-0245\\
2MASS J22573178-0651513 & 22 57 31.79 & -06 51 51.2 & 17.42 & 17.370 & 14.731 & 1.926 & 0.317 & T1 &\\
2MASS J23054834-0314121 & 23 05 48.29 & -03 14 11.7 & 18.59 & 17.810 & 14.175 & 2.600 & 0.318 & T1 &\\
2MASS J23062901-4951453 & 23 06 29.01 & -49 51 45.3 & 18.12 & 17.670 & 14.417 & 1.962 & 0.118 & T1 &\\
2MASS J23180937-1125452 & 23 18 09.36 & -11 25 45.5 & 17.49 & 17.230 & 14.745 & 1.866 & 0.042 & T1 &\\
2MASS J23253138-3010560 & 23 25 31.39 & -30 10 56.1 & 19.29 & 15.010 & 10.999 & 2.447 & 0.417 & PT1 &\\
2MASS J23305477-0956480 & 23 30 54.78 & -09 56 48.1 & 18.35 & 17.630 & 14.190 & 1.997 & 0.250 & T1 & 2MASX J23305473-0956479\\
2MASS J23342108-1421319 & 23 34 21.12 & -14 21 31.3 & 16.67 & 17.020 & 13.971 & 1.861 & 0.243 & T1 & 1RXS J233421.6-142128\\
2MASS J23352324-3945052 & 23 35 23.26 & -39 45 05.2 & 18.40 & 17.770 & 14.884 & 1.960 & 0.019 & T1 &\\
2MASS J23430906-5412369 & 23 43 09.09 & -54 12 36.8 & 17.81 & 18.230 & 14.609 & 2.189 & 0.624 & PT1 & SUMSS J234308-541238\\
2MASS J23441698-3322367 & 23 44 16.95 & -33 22 36.4 & 18.97 & 18.120 & 14.693 & 2.312 & 0.233 & T1 &\\
2MASS J23524661-2116467 & 23 52 46.60 & -21 16 46.9 & 17.30 & 17.120 & 14.751 & 1.937 & 0.280 & T1 & 1RXS J235247.1-211644\\
2MASS J23582823-2259320 & 23 58 28.22 & -22 59 31.9 & 18.09 & 16.230 & 13.415 & 2.006 & 0.114 & PT1 & 2MASX J23582823-2259322\\
\enddata
\tablenotetext{a}{In J2000.0, units are hours, minutes, seconds for R.A.,
            and degrees, arcminutes, arcseconds for Dec.}
\tablenotetext{b}{T1 = securely identified type 1 AGN;
            PT1 = probable identification}
\end{deluxetable}

\clearpage

\begin{deluxetable}{lcccccccl}
\tabletypesize{\scriptsize}
\rotate
\tablecaption{Type 2 AGN\tablenotemark{a}\label{type2agn}}
\tablewidth{0pt}
\tablehead{
\colhead{Name} &
\colhead{R.A.\tablenotemark{b}} & 
\colhead{Dec.\tablenotemark{b}} & 
\colhead{$b_J$} & 
\colhead{$r_F$} & 
\colhead{$K$} & 
\colhead{$J-K_s$} & 
\colhead{Redshift} &
\colhead{Other Name}
}
\startdata
2MASS J00041584-4938500 & 00 04 15.83 & -49 38 50.2 & 18.75 & 17.420 & 14.399 & 1.776 & 0.199 & 2MASX J00041588-4938507\\
2MASS J00112456-0808506 & 00 11 24.57 & -08 08 50.6 & 19.92 & 17.560 & 14.775 & 2.069 & 0.136 & APMUKS B000851.08-08252\\
2MASS J00153042-0750548 & 00 15 30.43 & -07 50 54.7 & 19.27 & 17.780 & 14.217 & 2.080 & 0.178 & 2MASX J00153041-0750549\\
2MASS J00213743-3726445 & 00 21 37.41 & -37 26 44.8 & 18.75 & 17.450 & 14.304 & 2.254 & 0.197 &\\
2MASS J00240550-1540006 & 00 24 05.51 & -15 40 00.4 & 19.07 & 17.680 & 13.926 & 2.099 & 0.241 & NVSS J002405-154004\\
2MASS J01023200-2151264 & 01 02 31.99 & -21 51 26.6 & 12.24 & 14.830 & 14.890 & 1.065 & 0.040 & 2MASX J01023199-2151267\\
2MASS J01220424-5355324 & 01 22 04.19 & -53 55 33.1 & 18.89 & 17.710 & 14.606 & 2.006 & 0.316 & APMUKS B012002.43-541113.6\\
2MASS J01231824-2207033 & 01 23 18.25 & -22 07 03.2 & 19.55 & 17.830 & 14.031 & 2.060 & 0.388 &\\
2MASS J01265229-4556045 & 01 26 52.28 & -45 56 04.5 & 19.77 & 17.810 & 14.903 & 1.508 & 0.294 & IRAS F01247-4611\\
2MASS J01291045-4331345 & 01 29 10.46 & -43 31 34.6 & 20.26 & 17.830 & 14.823 & 2.003 & 0.282 &\\
2MASS J01413624-1747177 & 01 41 36.25 & -17 47 17.6 & 17.83 & 17.260 & 14.477 & 1.795 & 0.311 &\\
2MASS J02171791-2638482 & 02 17 17.90 & -26 38 48.3 & 19.16 & 17.900 & 14.775 & 2.140 & 0.201 &\\
2MASS J02333333-0635397 & 02 33 33.33 & -06 35 39.5 & 19.10 & 17.570 & 14.403 & 2.051 & 0.185 &\\
2MASS J02340157-1934369 & 02 34 01.59 & -19 34 37.1 & 17.96 & 17.730 & 15.105 & 1.932 & 0.346 &\\
2MASS J02345773-3349442 & 02 34 57.73 & -33 49 44.4 & 18.37 & 16.580 & 14.802 & 1.756 & 0.117 & 2dFGRS S518Z006\\
2MASS J03143680-0557035 & 03 14 36.79 & -05 57 03.2 & 18.15 & 17.520 & 14.739 & 2.124 & 0.107 & APMUKS B031208.24-060810.4\\
2MASS J03232591-2110586 & 03 23 25.90 & -21 10 58.7 & 19.41 & 17.970 & 14.648 & 1.989 & 0.352 &\\
2MASS J03340366-1454428 & 03 34 03.66 & -14 54 43.0 & 18.63 & 17.670 & 13.928 & 1.967 & 0.167 & APMUKS B033144.22-150442.8\\
2MASS J03474693-2424458 & 03 47 46.90 & -24 24 45.8 & 18.59 & 18.160 & 15.214 & 1.674 & 0.272 &\\
2MASS J03535649-1605395 & 03 53 56.48 & -16 05 39.9 & 19.90 & 17.390 & 15.057 & 1.964 & 0.066 &\\
2MASS J04043951-0930307 & 04 04 39.52 & -09 30 30.7 & 18.75 & 17.160 & 14.952 & 2.402 & 0.359 &\\
2MASS J04082380-2152363 & 04 08 23.81 & -21 52 36.3 & 19.93 & 17.600 & 13.498 & 2.636 & 0.207 &\\
2MASS J04192642-0600032 & 04 19 26.42 & -06 00 03.3 & 19.06 & 17.880 & 14.601 & 1.952 & 0.223 &\\
2MASS J04221665-2315307 & 04 22 16.65 & -23 15 30.9 & 18.96 & 18.100 & 14.758 & 1.914 & 0.156 & APMUKS B042009.08-232228.6\\
2MASS J04242530-3604244 & 04 24 25.30 & -36 04 24.6 & 18.35 & 17.210 & 13.047 & 2.825 & 0.150 & IRAS F04226-3611\\
2MASS J04402018-0802175 & 04 40 20.18 & -08 02 17.4 & 19.32 & 17.960 & 13.916 & 2.070 & 0.189 & 2MASX J04402018-0802174\\
2MASS J04473869-4124594 & 04 47 38.70 & -41 24 59.4 & 19.04 & 17.380 & 13.550 & 2.003 & 0.226 & 2MASX J04473873-4124592\\
2MASS J04545129-3513006 & 04 54 51.28 & -35 13 00.9 & 19.29 & 17.860 & 14.256 & 2.208 & 0.240 &\\
2MASS J04562039-1701279 & 04 56 20.41 & -17 01 28.0 & 18.54 & 17.500 & 14.188 & 1.941 & 0.188 &\\
2MASS J04570851-2415069 & 04 57 08.53 & -24 15 06.7 & 19.30 & 17.760 & 14.445 & 2.518 & 0.360 &\\
2MASS J05003208-6032458 & 05 00 32.09 & -60 32 45.8 & 19.67 & 17.650 & 14.153 & 2.216 & 0.317 &\\
2MASS J05015803-2604343 & 05 01 58.04 & -26 04 34.2 & 17.21 & 16.630 & 14.043 & 2.007 & 0.037 & 2MASX J05015805-2604341\\
2MASS J05142233-2227103 & 05 14 22.32 & -22 27 10.4 & 18.41 & 17.930 & 15.160 & 1.966 & 0.357 &\\
2MASS J05302635-3452048 & 05 30 26.35 & -34 52 04.9 & 19.28 & 17.350 & 12.922 & 2.307 & 0.298 & NVSS J053026-345207\\
2MASS J05513207-5655295 & 05 51 32.04 & -56 55 29.5 & 19.17 & 17.660 & 13.908 & 2.140 & 0.257 & PKS 0550-569\\
2MASS J09490740-1211328 & 09 49 07.42 & -12 11 32.8 & 17.81 & 17.250 & 13.624 & 2.323 & 0.093 &\\
2MASS J10505072-1000060 & 10 50 50.71 & -10 00 05.9 & 18.53 & 17.570 & 14.223 & 2.123 & 0.193 & NVSS J105050-100001\\
2MASS J11072722-0125110 & 11 07 27.22 & -01 25 11.0 & 18.63 & 17.520 & 14.135 & 1.894 & 0.169 & 2MASX J11072722-0125117\\
2MASS J11575615-0453498 & 11 57 56.19 & -04 53 49.4 & 19.03 & 17.760 & 15.378 & 1.513 & 0.190 &\\
2MASS J12294408-1009001 & 12 29 44.07 & -10 09 00.2 & 19.71 & 17.950 & 14.224 & 2.323 & 0.268 & NVSS J122944-100900\\
2MASS J13021118-1201264 & 13 02 11.17 & -12 01 26.6 & 18.56 & 17.320 & 14.061 & 2.450 & 0.191 & 2MASX J13021117-1201265\\
2MASS J14144317-1556194 & 14 14 43.16 & -15 56 19.3 & 19.57 & 17.960 & 14.370 & 1.904 & 0.212 &\\
2MASS J14255845-1553274 & 14 25 58.46 & -15 53 27.5 & 19.37 & 17.940 & 14.377 & 2.235 & 0.284 &\\
2MASS J14495686-0301555 & 14 49 56.88 & -03 01 55.6 & 18.68 & 17.580 & 14.481 & 2.085 & 0.246 & LCRS B144720.8-024932\\
2MASS J15250421-0111569 & 15 25 04.23 & -01 11 57.2 & 18.66 & 17.750 & 14.799 & 2.048 & 0.123 & SDSS J152504.19-011156.5\\
2MASS J15320140-0858124 & 15 32 01.40 & -08 58 12.4 & 19.25 & 17.970 & 14.694 & 2.077 & 0.157 &\\
2MASS J20473227-4841084 & 20 47 32.27 & -48 41 08.5 & 18.32 & 16.920 & 13.682 & 2.163 & 0.145 & 2MASX J20473225-4841082\\
2MASS J21011126-2143329 & 21 01 11.26 & -21 43 33.1 & 18.21 & 17.530 & 13.698 & 2.065 & 0.304 &\\
2MASS J21291755-6238414 & 21 29 17.55 & -62 38 41.4 & 18.94 & 18.090 & 15.003 & 2.037 & 0.232 & APMUKS B212523.73-625150.2\\
2MASS J21300805-0155571 & 21 30 08.05 & -01 55 57.1 & 19.75 & 17.950 & 14.793 & 2.107 & 0.290 & RX J2130.2-0156\\
2MASS J21381991-1001522 & 21 38 19.93 & -10 01 52.2 & 18.72 & 17.530 & 14.174 & 1.958 & 0.206 & 2MASX J21381993-1001522\\
2MASS J21454002-2919369 & 21 45 40.02 & -29 19 36.9 & 19.44 & 17.410 & 14.094 & 2.009 & 0.341 & APMUKS B214245.16-293329.4\\
2MASS J21462665-4212558 & 21 46 26.65 & -42 12 56.1 & 18.96 & 17.830 & 15.181 & 1.877 & 0.209 &\\
2MASS J22024635-1055517 & 22 02 46.36 & -10 55 51.0 & 18.83 & 17.430 & 12.847 & 2.526 & 0.238 & APMUKS B220006.09-111022.8\\
2MASS J22060976-3558119 & 22 06 09.70 & -35 58 12.1 & 20.20 & 17.980 & 14.999 & 1.702 & 0.056 &\\
2MASS J23200024-4309095 & 23 20 00.23 & -43 09 09.5 & 19.01 & 17.670 & 14.096 & 2.054 & 0.177 & 2MASX J23200025-4309100\\
2MASS J23572052-5503091 & 23 57 20.53 & -55 03 09.3 & 19.23 & 17.720 & 14.809 & 1.989 & 0.300 &\\
\enddata
\tablenotetext{a}{All these are {\it probable} identifications (See section 4.1)}
\tablenotetext{b}{In J2000.0, units are hours, minutes, seconds for R.A.,
            and degrees, arcminutes, arcseconds for Dec.}
\end{deluxetable}

\end{document}